\documentclass[twocolumn,preprintnumbers,amsmath,amssymb]{revtex4}
\usepackage{amsmath}
\usepackage{extarrows}
\usepackage{stmaryrd}
\usepackage[ruled]{algorithm2e}
\usepackage{graphicx}
\usepackage{dcolumn}
\usepackage{bm}
\usepackage[all]{xy}
\usepackage{indentfirst}
\usepackage{bbding}
\usepackage{color}
\usepackage{appendix}
\usepackage{pifont}
\usepackage{amsmath}
\usepackage{multirow}
\usepackage{mathrsfs}
\usepackage{bbm}
\usepackage{euscript}
\usepackage{amssymb}
\begin{document}
\title{Robust Multipartite Entanglement Without Entanglement Breaking}

\author{Ming-Xing Luo$^{1}$, Shao-Ming Fei$^{2,3}$}

\affiliation{\small{}1. School of Information Science and Technology, Southwest Jiaotong University, Chengdu 610031, China;
\\
2. School of Mathematical Sciences, Capital Normal University, Beijing 100048, China;
\\
3. Max-Planck-Institute for Mathematics in the Sciences, 04103 Leipzig, Germany}

\begin{abstract}
Entangled systems in experiments may be lost or offline in distributed quantum information processing. This inspires a general problem to characterize quantum operations which result in breaking of entanglement or not. Our goal in this work is to solve this problem both in single entanglement and network scenarios. We firstly propose a local model for characterizing all entangled states that are breaking for losing particles. This implies a simple criterion for witnessing single entanglement such as generalized GHZ states and Dicke states. It further provides an efficient witness for entangled quantum networks depending on its connectivity such as $k$-independent quantum networks, completely connected quantum networks, and $k$-connected quantum networks. These networks are universal resources for measurement-based quantum computations. The strong nonlocality can be finally verified by using nonlinear inequalities. These results show distinctive features of both single entangled systems and entangled quantum networks.
\end{abstract}
\maketitle

\section{Introduction}

As one of the remarkable features in quantum mechanics, quantum entanglement has attracted great attentions \cite{EPR}. The quantum correlations generated by local measurements on entangled two-spin systems can not be reproduced from classical physics. These nonlocal quantum correlations are verified by violating Bell inequalities \cite{Bell,CHSH,Gis}. For multipartite scenarios the quantum correlations may be verified in specific local models under different assumptions \cite{GHZ,Merm,Sy,BCPS}. The entangled states have become important resources in various ongoing studies \cite{LJL,ZZH,HHH}.

Experimentally states based on atomic ensembles provide attractive systems for both the storage of quantum information and the coherent conversion of quantum information between atomic and optical degrees of freedom \cite{HSS}. However, under the evolution $U_t=e^{it\mathbf{H}}$ with Hamiltonian $\mathbf{H}$, the nucleus of an unstable isotope may lose one of several particles including neutrons, alpha particles, electrons or positrons \cite{PKGR}, see Fig. 1(a). The particle-lose channel ${\cal E}_{S}={\rm Tr}_S$, where the partial trace goes over the lost particles contained in set $S$, may be an entanglement-breaking channel \cite{Holevo,HSR}. The output is given by $\rho_{out}={\cal E}_{S}(\rho_{in})$ with input $\rho_{in}$. It is marvelous that some states like Dicke states keep entangled after passing through the particle-lose channel \cite{GZ,QABZ,BPB,SZD}. This inspires a natural problem of characterizing multipartite quantum systems in term of particle-lose channels.

Compared with entangled qubit systems \cite{EPR,GHZ}, high-dimensional entangled systems may inherent local tensor decompositions, see Fig.1 (b), which intrigue distributed experiments in preparing quantum networks \cite{Kim,LJL}. One example is the cluster states that are universal resources for measurement-based quantum computations \cite{Cluster}. In this case, some experimental devices may be not online in large-scale or remote tasks, regarded as party-lose noises in which all the local particles shared by one party are taken as one unavailable high-dimensional particle or quantum sources. Different from the permutationally symmetric states \cite{GZ,QABZ,BPB,SZD}, these entangled systems depend on network configurations, which give rise to an interesting problem in characterizing distributed quantum resources for quantum information processing.

\begin{figure}
\begin{center}
\resizebox{210pt}{120pt}{\includegraphics{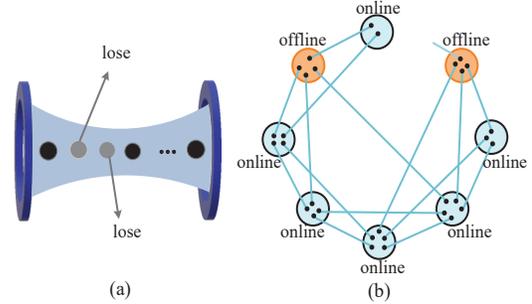}}
\end{center}
\caption{\small (Color online) (a) Atomic ensembles with losing particles. (b) Quantum networks with offline parties. High-dimensional states may be prepared by distributed experiments with small entangled systems such as Affleck-Kennedy-Lieb-Tasaki (AKLT) system \cite{AKLT}.}
\label{fig-1}
\end{figure}

In this work, we explore the multipartite entangled systems both in Bell and network scenarios. The main idea is to characterize the strong correlations against particle loss. We firstly propose a local model for featuring fragile entangled states, that is, the entanglement breaks for some particle-lose channels. This model is stronger than the biseparable model \cite{Sy} because GHZ states \cite{GHZ} are not entangled in the present model. It is also different from the network local model \cite{NWR,Kraft,Luo2020} that can witness GHZ states \cite{GHZ}. Especially, the generalized GHZ states are the unique kind of fragile entangled states that become fully separable under any one  particle-loss channel. Instead, the generalized Dicke states \cite{Toth} keep entangled after particle loss. This shows distinctive features beyond previous models \cite{Sy,GHZ,Cluster,SAS,GS}. Moreover, it is extended to witness distributed resources such as $k$-independent quantum networks \cite{Luo2018}, completely connected quantum networks, or $k$-connected quantum networks \cite{AKLT,Cluster} according to the connectivity. The strong nonlocality is verified by using nonlinear Bell inequalities. The results may shed new light on network entanglement and quantum information processing such as distributed quantum computation.

\section{Fragile multipartite entanglement}

Let $\mathcal{H}_{A_j}$ be the Hilbert space associated with the particle ${A_j}$, and $\rho_{A_1\cdots A_n}$ an $n$-partite state in $\otimes_{j=1}^n\mathcal{H}_{A_j}$. Denote $S$ as a subset of $\{A_1,\cdots, A_n\}$, and $\overline{S}$ as the complement set of $S$. Let ${\cal E}_{S}(\cdot{})$ be a complete positive trace preserving (CPTP) mapping of the particle-lose channel associated with the particle set $S$, that is, ${\cal E}_{S}(\rho)=\sum_{j\in K_S}E_{j}\rho{}E^\dag_j$ for any state $\rho$, where $K_S$ denotes the Kraus operator decomposition of ${\cal E}_{S}(\cdot{})={\rm Tr}_{A_j\in S}(\cdot{})$ and ${\rm Tr}_{A_j\in S}(\cdot{})$ denotes the partial trace operator of particles in $S$. For instance, when $S=\{A_1\}$, we have  $E_j=\langle j|_{A_1}\otimes \mathbbm{1}_{A_2\cdots A_n}$, where $\mathbbm{1}_{A_2\cdots A_n}$ is the identity operator on particles $A_2, \cdots, A_n$. Let $\rho_{(S)}$ denote the output state after $\rho$ passing through the particle-lose channel ${\cal E}_{S}(\cdot{})$, that is, $\rho_{(S)}={\cal E}_{S}(\rho)$. The main motivation here is to explore new features of multipartite entangled systems both in single source and network configurations, which cannot be carried out by using the biseparable model \cite{Sy}, network Bell inequalities \cite{Luo2018} or network local model \cite{NWR,Kraft,Luo2020}.

Our main result in what follows is based on the biseparable model \cite{Sy}, that is, an $n$-particle state of $\rho_{A_1\cdots A_n}$ on Hilbert space $\otimes_{j=1}^n\mathcal{H}_{A_j}$ is genuinely $n$-partite entanglement if it cannot be decomposed into as
\begin{eqnarray}
\rho_{bs}=\sum_{S}\sum_jp_{j;S}\rho_{j}^{(S)}\otimes\rho_{j}^{(\overline{S})}
\label{eqn0}
\end{eqnarray}
where $S$ and $\overline{S}$ are any bipartition of $\{A_1, \cdots, A_n\}$, $\{p_{j;S}\}$ is a probability distribution, $\rho_{j}^{(S)}$ and $\rho_{j}^{(\overline{S})}$ are respectively states of particles in $S$ and $\overline{S}$. Here, $\rho_{bs}$ in Eq.(\ref{eqn0}) is named as biseparable state \cite{Sy}.

\textbf{Definition 1}. An $n$-partite state $\rho$ is particle-lose separable if $\rho_{(S)}$ is biseparable for some set $S$ with no more than $n-2$ particles. Otherwise, $\rho$ is robust entanglement.

Consider the GHZ state $|\Phi\rangle=\frac{1}{\sqrt{2}}(|000\rangle+|111\rangle)$ \cite{GHZ}, the reduced state is given by $\rho=\frac{1}{2}(|00\rangle\langle 00|+|11\rangle\langle 11|)$ which is separable. This implies the GHZ state is particle-lose separable. Instead, for Dicke state of $|\Phi\rangle=\frac{1}{\sqrt{3}}(|001\rangle+|010\rangle+|100\rangle)$ \cite{Toth}, we will prove it is robust entanglement in the next section. This means that Definition 1 gives stronger conditions than the $k$-uniform entanglement \cite{GHZ} because all the $k$-uniform $n$-particle entangled states which require all its reductions to $k$-qudits being maximally mixed are particle-lose separable for $k\leq n-2$. Moreover, different from the lose channel with fixed particles for symmetric Dicke states \cite{BPB,SZD} of general symmetric states, the present model may lose any particles with the only assumption of particle number.

\subsection{Qubit states}

Consider the generalized GHZ state \cite{GHZ,Ghose09} given by
\begin{eqnarray}
|GHZ\rangle_{A_1\cdots A_n}=\cos\theta|0\rangle^{\otimes n}+\sin\theta|1\rangle^{\otimes n},
\label{eqnaa1}
\end{eqnarray}
where $\theta\in (0,\frac{\pi}{4})$. Note that $|GHZ\rangle$ is genuinely multipartite entangled \cite{Sy} for $\theta\in (0,\frac{\pi}{4})$. However, it reduces to a fully separable state after losing any one qubit. Namely, it is weakly entangled with respect to the particle-loss noises. Interestingly, this kind of states is unique under local unitary operations.

\textbf{Proposition 1}. For a genuinely pure entangled qubit state $|\Phi\rangle_{A_1\cdots A_n}$ on Hilbert space $\otimes_{j=1}^n\mathbbm{H}_{A_j}$, it is equivalent to the generalized GHZ state in Eq.(\ref{eqnaa1}) if and only if the reduced state of $\rho_{(A_{j})}={\cal{} E}_{A_j}(|\Phi\rangle\langle \Phi|)$ is fully separable for any $j\in\{1, \cdots, n\}$, that is, there are local unitary operations $U_j$ such that
\begin{eqnarray}
(\otimes_{j=1}^nU_j)|\Phi\rangle_{A_1\cdots A_n}=|GHZ\rangle_{A_1\cdots A_n}.
\label{eqn4}
\end{eqnarray}

\textbf{Proof}. For a generalized GHZ state in Eq.(\ref{eqnaa1}), it is easy to prove that $\rho_{(A_{j})}=\cos^2\theta(|0\rangle\langle0|)^{\otimes n-1}+\sin^2\theta(|1\rangle\langle 1|)^{\otimes n-1}$, which is fully separable. This has proved the sufficient condition.

In what follows, we prove the necessary condition, that is, for a genuinely pure entangled qubit state $|\Phi\rangle_{A_1\cdots A_n}$ on Hilbert space $\otimes_{j=1}^n\mathbbm{H}_{A_j}$, if the reduced state of $\rho_{(A_{j})}={\cal{} E}_{A_j}(|\Phi\rangle\langle \Phi|)$ is fully separable for any $j\in\{1, \cdots, n\}$ there are local unitary operations $U_j$ satisfying Eq.(\ref{eqn4}). From this assumption, we get $|\Phi\rangle_{A_1\cdots A_n}$ is a genuinely $n$-partite entanglement while $\rho_{(A_j)}$ is fully separable.

The followed proof is completed by induction on $j$.

For $j=1$, we consider the bipartition $\{A_1\}$ and $\{A_2,\cdots, A_n\}$ of $\{A_1$, $\cdots$, $A_n\}$. The Schmidt decomposition of $|\Phi\rangle$ is given by
\begin{eqnarray}
|\Phi\rangle_{A_1\cdots A_n}=\sum_{i=0}^1\sqrt{p_i}|\psi_i\rangle_{A_1}
|\Phi_i\rangle_{A_2\cdots A_n},
\label{D2}
\end{eqnarray}
where $\{|\psi_i\rangle, \forall i\}$ are orthogonal states of qubit $A_1$, and $\{|\Phi_i\rangle, \forall i\}$ are orthogonal states of qubits $A_2, \cdots, A_n$, and $p_i$ are positive Schmidt coefficients.

For $j=2$, we will prove that there two qubits $A_1$ and $A_2$ are symmetric under local unitary operations, that is, there are unitary operations $U_{A_t}^{(t)}$ on the qubits $A_1$ and $A_2$ satisfying
\begin{eqnarray}
(U_{A_1}^{(1)}\otimes{}U_{A_2}^{(2)}\otimes \mathbbm{1}_3)|\Phi\rangle=\sum_{s=0}^1\sqrt{p_s}|ss\rangle_{A_1A_2}
|\Psi_s\rangle_{A_3\cdots A_n},
\label{D222}
\end{eqnarray}
where $\mathbbm{1}_{3}$ denotes the identity operator on qubits $A_3, \cdots{}, A_n$, and $|\Psi_s\rangle_{A_3\cdots A_n}$ are orthogonal states on qubits $A_3, \cdots{}, A_n$. In fact, from the assumption of necessary condition, ${\cal{} E}_{A_1}(|\Phi\rangle\langle \Phi|)$ is fully separable. Combined with Eq.(\ref{D2}) the reduced state of $\varrho_{(A_1)}$ is given by
\begin{eqnarray}
\varrho_{(A_1)}
&=&{\cal{} E}_{A_1}(|\Phi\rangle\langle \Phi|)
\nonumber\\
&=&{\cal{} E}_{A_1}((U_{A_1}^{(1)}\otimes\mathbbm{1}_2)|\Phi\rangle\langle \Phi|((U_{A_1}^{(1)})^\dag\otimes\mathbbm{1}_2))
\label{D3}
\\
&=&\sum_{s=0}^1\langle \psi_s|_{A_1}|\Phi\rangle\langle \Phi|\psi_s\rangle_{A_1}
\nonumber\\
&=&\sum_{s=0}^1p_s|\Phi_s\rangle_{A_2\cdots{}A_n}\langle \Phi_s|,
\label{D4}
\end{eqnarray}
where $\mathbbm{1}_2$ is the identity operator on qubits $A_2, \cdots, A_n$. Eq.(\ref{D3}) is followed from the invariance of the particle-lose channel ${\cal{} E}_{A_1}(\cdot)$ under local unitary operations in the orthogonal states $|\psi_s\rangle_{A_1}$. Another method to derive Eq.(\ref{D4}) is using the normal form of multipartite pure entanglement under local unitary operations \cite{Kraus2009}, that is, one firstly performs the unitary operation $U_{A_1}^{(1)}$ on $|\Phi\rangle$ in Eq.(\ref{D2}) and then implements the particle-lose channel ${\cal{} E}_{A_1}(\cdot)$.

From the assumption of necessary condition, $\varrho_{(A_1)}$ in Eq.(\ref{D4}) is fully separable. So, there are four cases for $|\Phi_0\rangle$ and $|\Phi_1\rangle$:
\begin{itemize}
\item[(C1)] Both states of $|\Phi_0\rangle$ and $|\Phi_1\rangle$ given in Eq.(\ref{D4}) are fully separable.
\item[(C2)] Only one state $|\Phi_0\rangle$ (for example) given in Eq.(\ref{D4}) is fully separable.
\item[(C3)] Both states of $|\Phi_0\rangle$ and $|\Phi_1\rangle$ given in Eq.(\ref{D4}) are entangled.
\item[(C4)] One state $|\Phi_0\rangle$ (for example) in Eq.(\ref{D4}) is biseparable \cite{Sy}.
\end{itemize}
The case (C1) will be proved in Lemma 1 shown in Appendix A. The case (C2) will be proved in Lemma 2 shown in Appendix B. The case (C3) will be proved in Lemma 3 shown in Appendix C. Moreover, the case (C4) will be reduced to the case (C2) in the following paragraph.

\textbf{Lemma 1}. If both states of $|\Phi_0\rangle$ and $|\Phi_1\rangle$ given in Eq.(\ref{D4}) are fully separable, then $|\Phi\rangle$ in Eq.(\ref{D2}) is equivalent to the generalized GHZ state in Eq.(\ref{eqnaa1}) under local unitary operations.

\textbf{Lemma 2}. If one state $|\Phi_0\rangle$ (for example) given in Eq.(\ref{D4}) is fully separable, then $|\Phi\rangle$ in Eq.(\ref{D2}) is equivalent to the generalized GHZ state in Eq.(\ref{eqnaa1}) under local unitary operations.

\textbf{Lemma 3}. If both states of $|\Phi_0\rangle$ and $|\Phi_1\rangle$ given in Eq.(\ref{D4}) are genuinely entangled, then the qubits $A_1$ and $A_2$ in $|\Phi\rangle$  in Eq.(\ref{D2}) are symmetric under local unitary operations.

Now, continue the proof. For the case (C1), from Lemma 1 $|\Phi\rangle$ in Eq.(\ref{D2}) is equivalent to the generalized GHZ state in Eq.(\ref{eqnaa1}) under local unitary operations, that is, all the qubits are symmetric under specific local unitary operations. For the case (C2), from Lemma 2 the state $|\Phi\rangle$ in Eq.(\ref{D2}) is equivalent to the generalized GHZ state in Eq.(\ref{eqnaa1}) under local unitary operations. This has completed the proof. In both cases we have proved the necessary condition of Proposition 1.

For the case (C3), from Lemma 3 there are two qubits $A_1$ and $A_2$ of $|\Phi\rangle$ in Eq.(\ref{D2}) are symmetric under local unitary operations. This has proved Eq.(\ref{D222}).

In what follows, we prove the case (C4) will be reduced to the case (C3), that is, assume that $|\Phi_0\rangle_{A_2\cdots A_n}$ given in Eq.(\ref{D4}) is biseparable for one bipartition $\{A_2, \cdots, A_k\}$ and $\{A_{k+1},\cdots, A_n\}$ of $\{A_2,\cdots, A_n\}$ (for simplicity) \cite{Sy}, that is,
\begin{eqnarray}
|\Phi_0\rangle_{A_2\cdots A_n}=|\Phi_{00}\rangle_{A_2\cdots A_k}|\Phi_{01}\rangle_{A_{k+1}\cdots{}A_n},
\label{DDd1}
\end{eqnarray}
where $|\Phi_{00}\rangle$ is an genuinely entangled state of qubits $A_2,\cdots, A_k$, and $|\Phi_{01}\rangle$ is a general state (which may be not entangled) of qubits $A_{k+1},\cdots, A_n$. In this case, $|\Phi_1\rangle$ given in Eq.(\ref{D4}) is biseparable in terms of the same bipartition, that is,
\begin{eqnarray}
|\Phi_1\rangle_{A_2\cdots A_n}=|\Phi_{10}\rangle_{A_2\cdots A_k}|\Phi_{11}\rangle_{A_{k+1}\cdots{}A_n},
\label{DDd2}
\end{eqnarray}
where $|\Phi_{10}\rangle$ is a general state of qubits $A_2,\cdots, A_k$, and $|\Phi_{11}\rangle$ is a general state of qubits $A_{k+1},\cdots, A_n$. Otherwise, $|\Phi_1\rangle_{A_2\cdots A_n}$ given in Eq.(\ref{D4}) is a genuinely $n-1$-partite entanglement in the biseparable model \cite{Sy}. This implies that $\varrho_{(A_1)}$ in Eq.(\ref{D4}) is a genuinely $n-1$-partite entanglement from the definition of the biseparable model \cite{Sy}, that is, $\varrho_{(A_1)}$ is a mixed state of one biseparable state $|\Phi_0\rangle_{A_2\cdots A_n}$ and one genuinely $n-1$-partite entanglement $|\Phi_1\rangle_{A_2\cdots A_n}$, which cannot be decomposed into a mixture of all the biseparable states. This contradicts to the assumption that $\varrho_{(A_1)}$ is fully separable. Hence, $|\Phi_1\rangle_{A_2\cdots A_n}$ has the decomposition in Eq.(\ref{DDd2}).

From Eqs.(\ref{DDd1}) and (\ref{DDd2}), the reduced state of $\rho_{(A_1A_{k+1}\cdots A_n)}$ after $\rho_{(A_1)}$ in Eq.(\ref{D4}) passing through the particle-lose channel ${\cal E}_{A_{k+1}\cdots A_n}(\cdot)$ has the following decomposition as
\begin{eqnarray}
\rho_{(A_1A_{k+1}\cdots A_n)}&=&{\cal E}_{A_{k+1}\cdots A_n}(\rho_{(A_1)})
\nonumber\\
&:=&\sum_{s=0}^1q_s|\Phi_{s0}\rangle_{A_2\cdots A_k}\langle \Phi_{s0}|,
\label{Dd44}
\end{eqnarray}
where $\{q_s\}$ is a probability distribution which may be different from $\{p_s\}$ in Eq.(\ref{DDd2}). From the separability assumption of $\rho_{(A_1)}$, $\rho_{(A_1A_{k+1}\cdots A_n)}$ should be fully separable. Moreover, from assumption in Eq.(\ref{DDd1}), $|\Phi_{00}\rangle_{A_2\cdots A_k}$ is an genuinely entangled state of qubits $A_2, \cdots, A_k$. It follows that $|\Phi_{10}\rangle_{A_2\cdots A_k}$ should be entangled. Otherwise, that is, $|\Phi_{10}\rangle_{A_2\cdots A_k}$ is biseparable. This follows an entangled state $\rho_{(A_1A_{k+1}\cdots A_n)}$ in Eq.(\ref{Dd44}) because it is a mixed state of the genuinely entangled state $|\Phi_{00}\rangle_{A_2\cdots A_k}$ and the biseparable state $|\Phi_{10}\rangle_{A_2\cdots A_k}$. It contradicts to the assumption of the necessary condition that states $\rho_{(A_1A_{k+1}\cdots A_n)}$ should be fully separable. Hence, $\rho_{(A_1A_{k+1}\cdots A_n)}$ in Eq.(\ref{Dd44}) is the mixed state of two genuinely entangled states. From Lemma 3, three are two qubits $A_1$ and $A_v$ (with $v=2$ for simplicity) in $\rho_{(A_1A_{k+1}\cdots A_n)}$ in Eq.(\ref{Dd44}) are symmetric under local unitary operations, that is, two qubits $A_1$ and $A_2$ in $|\Phi\rangle_{A_1\cdots A_n}$ in Eq.(\ref{D2}) are symmetric under specific local unitary operations. We have completed the proof for the case (C4).

For any $j=k-1$, assume that qubits $A_{1}, \cdots, A_j$ are symmetric under local operations, that is, there are unitary operations $U_{A_{t}}^{(t)}$ on qubits $A_{t}$ such that
\begin{eqnarray}
&&(\otimes_{t=1}^{k-1}U_{A_t}^{(t)}\otimes \mathbbm{1}_{k})|\Phi\rangle_{A_{1}\cdots{}A_n}
\nonumber\\
&=&\sum_{s=0}^1\sqrt{p_s}|s\cdots {}s\rangle_{A_1\cdots A_{k-1}}
|\hat{\Phi}_s\rangle_{A_{k}\cdots A_n},
\label{D222sa}
\end{eqnarray}
where $\mathbbm{1}_{k}$ is the identity operator on qubits $A_{k}, \cdots{}, A_n$, and $|\hat{\Phi}_s\rangle_{A_{k}\cdots A_n}$ are orthogonal states. We will prove the result for $j=k$, that is, there is at least one qubit $A_{\ell_k}$ such that $A_1, \cdots, A_{k-1}$ and $A_{\ell_k}$ are symmetric under specific local unitary operations. In fact, from Eq.(\ref{D222sa}) we get the reduced state
\begin{eqnarray}
\rho_{(A_1\cdots{}A_{k-1})}
&=&{\cal{} E}_{A_1\cdots{}A_{k-1}}(|\Phi\rangle\langle \Phi|)
\nonumber\\
&=&\sum_{s=0}^1p_s|\hat{\Phi}_s\rangle_{A_{k}\cdots A_n}\langle \hat{\Phi}_s|
\label{D222ss}
\end{eqnarray}
The followed proof is similar to the proof after Eq.(\ref{D4}) by considering the four cases (C1)-(C4) for two states $|\hat{\Phi}_0\rangle_{A_{k}\cdots A_n}$ and $|\hat{\Phi}_1\rangle_{A_{j}\cdots A_n}$. Hence, there is at least one qubit $A_{\ell_k}$ ($\ell_k=k$ for simplicity) such that $A_1, \cdots, A_{k-1}$ and $A_{\ell_k}$ are symmetric under specific local unitary operations, that is, there are unitary operations $U_{A_{t}}^{(t)}$ on qubit $A_{t}$ such that
\begin{eqnarray}
&&(\otimes_{t=1}^{k}U_{A_t}^{(t)}\otimes \mathbbm{1}_{k+1})|\Phi\rangle
\nonumber
\\
&=&\sum_{s=0}^1\sqrt{p_s}|s\cdots {}s\rangle_{A_1\cdots A_k}
|\tilde{\Phi}_s\rangle_{A_{k+1}\cdots A_n},
\label{D222st}
\end{eqnarray}
where $\mathbbm{1}_{k+1}$ is the identity operator on qubits $A_{k+1}, \cdots{}, A_n$, and $|\tilde{\Phi}_s\rangle_{A_{k+1}\cdots A_n}$ are orthogonal states. This has completed the proof for $j=k$.

This procedure will be ended when $j=n-1$. Finally, all the qubits $A_1, \cdots, A_n$ are symmetric under specific local unitary operations. From Eq.(\ref{D222st}), $|\Phi\rangle$ in Eq.(\ref{D2}) is equivalent to the generalized $n$-qubit GHZ state in Eq.(\ref{eqnaa1}) under local unitary operations. This completes the proof. $\Box$ 

Proposition 1 implies the uniqueness of such weakly entangled qubit systems. It shows one difference between the present model and the biseparable model \cite{Sy} or the network model \cite{NWR,Kraft,Luo2020} in which the generalized GHZ states are entangled. Such property may hold for high-dimensional entangled pure states including the absolute maximal entangled states \cite{GW,HCL,HGS}, or the symmetric states which will be proved in the next section.

\subsection{High-dimensional entangled symmetric pure states}

In this subsection, we extend Proposition 1 to high-dimensional permutationally symmetric states. For an $n$-partite $d$-dimensional state $|\Phi\rangle_{A_1\cdots A_n}$ on Hilbert space $\otimes_{j=1}^n\mathbb{H}_{A_j}$, it is permutationally symmetric if $|\Phi\rangle_{A_1\cdots A_n}$ is invariant under any permutation operation (swapping the particles). Define a $d$-dimensional generalized GHZ state as
\begin{eqnarray}
|GHZ_d\rangle=\sum_{j=0}^{d-1}\alpha_j|j\cdots j\rangle_{A_1\cdots A_n},
\label{DDD1}
\end{eqnarray}
where $\alpha_j$'s satisfy $\sum_{j=0}^{d-1}\alpha_j^2=1$ and $\theta\in (0, \frac{\pi}{4}]$.

\textbf{Proposition 2}. For a permutationally symmetric $n$-partite entanglement $|\Phi\rangle_{A_1\cdots A_n}$, it is equivalent to the generalized GHZ state in Eq.(\ref{DDD1}) if and only if $\rho_{(A_{j})}={\cal{} E}_{A_j}(|\Phi\rangle\langle \Phi|)$ is fully separable for any $j\in \{1, \cdots, n\}$, that is, there are local unitary operations $U_{A_j}^{(j)}$ on particle $A_j$ such that
\begin{eqnarray}
(\otimes_{j=1}^nU_{A_j}^{(j)})|\Phi\rangle_{A_1\cdots A_n}=|GHZ_d\rangle.
\label{eqnDD2}
\end{eqnarray}

\textbf{Proof}. The sufficient condition is easily followed from Eq.(\ref{DDD1}). In what follows, we prove that a permutationally symmetric $n$-partite entanglement $|\Phi\rangle$ is equivalent to the generalized GHZ state in Eq.(\ref{DDD1}) if $\rho_{(A_{j})}={\cal{} E}_{A_j}(|\Phi\rangle\langle \Phi|)$ is fully separable for any $j\in \{1, \cdots, n\}$. Note that $\{|GHZ_d\rangle, |D_{k,n}\rangle, \forall k\}$ consists the orthogonal basis of permutationally symmetric subspace of $\otimes_{j=1}^n\mathbb{H}_{A_j}$, where $|D_{k,n}\rangle$ denotes the maximally entangled $n$-particle Dicke states \cite{Toth}. It means that any permutationally symmetric state $|\Phi_{sy}\rangle_{A_1\cdots A_n}$ can be decomposed into
\begin{eqnarray}
|\Phi_{sy}\rangle=\sum_{k}\beta_k|D_{k,n}\rangle_{A_1\cdots A_n}+\beta|GHZ_d\rangle_{A_1\cdots A_n}
\label{DDD2}
\end{eqnarray}
where $\beta_k$'s and $\beta$ satisfy $\sum_i|\beta_k|^2+\beta^2=1$.

Note that
\begin{eqnarray}
&&\sum_{k}\beta_k|D_{k,n}\rangle_{A_1\cdots A_n}
\nonumber\\
&=&
\sum_{j_1=0}^{d-1}|j_{1}\rangle_{A_1}\sum_{k}\frac{\beta_k}{\sqrt{N_{k-j_1,n}}}
|D_{k-j_1}\rangle_{A_2\cdots A_n},
\label{DDD3}
\end{eqnarray}
where $|D_{k-j_1,n}\rangle$ denotes $n-1$-particle Dicke state with excitations $k-j_{1}$ defined by
\begin{eqnarray}
|D_{k-j_1,n}\rangle=\frac{1}{\sqrt{N_{k-j_{1},n}}}\sum_{j_2+\cdots +j_n=k-j_{1}}|j_2\cdots{}j_n\rangle
\label{DDD3a}
\end{eqnarray}
and $N_{k-j_{1},n}$ denotes the normalization constant of $|D_{k-j_1,n}\rangle$.

The reduced state after $|\Phi_{sy}\rangle$ in Eq.(\ref{DDD2}) passing through the particle-lose channel ${\cal{} E}_{A_1}(\cdot)$ is given by
\begin{eqnarray}
\rho_{(A_1)}&=&{{\cal{} E}}_{A_1}(|\Phi_{sy}\rangle\langle \Phi_{sy}|)
\nonumber
\\
&=&\beta^2\sum_{j=0}^{d-1}\alpha_j^2|j\cdots j\rangle_{A_2\cdots{}A_n}\langle j\cdots j|
 \nonumber
 \\
 &&
+{\cal{} E}_{A_1}(\sum_{k,s}\beta_k\beta_s|D_{k,n}\rangle_{A_2\cdots{}A_n}\langle D_{s,n}|)
 \nonumber
 \\
 &=&\sum_{k}\frac{\beta_k^2}{N_{k-j_1,n}}|D_{k-j_1,n-1}\rangle_{A_2\cdots{}A_n}\langle D_{k-j_1,n-1}|
  \nonumber
 \\
 &&+
  \beta^2\sum_{j=0}^{d-1}\alpha_j^2|j\cdots j\rangle_{A_2\cdots{}A_n}\langle j\cdots j|
 \label{DDD4}
\end{eqnarray}
From Eq.(\ref{DDD3a}) the states $|D_{k_1-j_1,n-1}\rangle$ and $|D_{k_2-j_1,n-1}\rangle$ have different excitations for any $k_1\not=k_2$. This implies that all the states of $\{|D_{k-j_1,n-1}\rangle, \forall k\}$ are defined on different orthogonal subspaces ${\cal H}_{k_1}$ and ${\cal H}_{k_2}$ of $\otimes_{j=2}^n\mathbb{H}_{A_j}$, where ${\cal H}_{k_s}$ is spanned by all the states $|j_2\cdots{}j_n\rangle$ with $j_2+\cdots +j_n=k_s-j_{1}$, $s=1,2$. Hence, $\{|D_{k-j_1,n-1}\rangle, \forall k\}$ are orthogonal states. Moreover, the state of $|D_{k-j_1,n-1}\rangle$ is a genuinely entangled \cite{Sy}. It follows that
\begin{eqnarray}
|\Psi\rangle_{A_2\cdots{}A_n}=\sum_{k}\frac{\beta_k^2}{N_{k-j_1,n}}|D_{k-j_1,n-1}\rangle\langle D_{k-j_1,n-1}|
\end{eqnarray}
is genuinely entangled \cite{DPR,Dicke4} if $\sum_{k}\beta_k^2\not=0$. Moreover, from the assumption, $\rho_{(A_1)}$ is fully separable. Note that $\sum_{j=0}^{d-1}\alpha_j^2|j\cdots j\rangle_{A_2\cdots{}A_n}\langle j\cdots j|$ is fully separable. This implies that $\sum_{k}\beta_k^2=0$, that is, $\beta_k=0$ for all $k$'s. It follows from Eq.(\ref{DDD4}) that
 \begin{eqnarray}
\rho_{(A_1)}=\beta^2\sum_{j=0}^{d-1}\alpha_j^2|j\cdots j\rangle_{A_2\cdots{}A_n}\langle j\cdots j|
\label{DDD5}
\end{eqnarray}
Hence, from Eq.(\ref{DDD5}) and the symmetry assumption, it follows that $|\Phi_{sy}\rangle_{A_1\cdots A_n}$ in Eq.(\ref{DDD2}) is equivalent to a generalized GHZ state $|GHZ_d\rangle$ defined in Eq.(\ref{DDD1}). This completes the proof. $\Box$

Proposition 2 has considered the permutationally symmetric $n$-partite entangled states. A natural problem is to consider general high-dimensional entangled states. One example is  $|\Phi\rangle_{A_1A_2A_3}=\frac{1}{2}(|000\rangle+|011\rangle+|120\rangle+|131\rangle)$. The reduced state after $|\Phi\rangle_{A_1A_2A_3}$ passing through the particle-lose channel ${\cal E}_{A_2}(\cdot)$ is separable but not for other channels ${\cal E}_{A_1}(\cdot)$ and ${\cal E}_{A_3}(\cdot)$. This means that the reduced states after passing particle-losing channel have different properties. Hence, it may be difficult to get general result for all high-dimensional entangled states.

\section{Robust multipartite entanglement}

Let ${\cal B}_d$ consist of all the particle-lose separable states. For any state $\rho\not\in {\cal B}_d$, Definition 1 shows strong robustness in terms of particle-lose channels. One example is W state \cite{DVC,AR,Sy}, which can be witnessed by using the PPT criterion \cite{PPT}. Different from entanglement theories \cite{Bell,Sy,Luo2020}, ${\cal B}_d$ is only star-convex, see Appendix D. The star-convex means that there is one center state $\rho_0\in {\cal B}_d$ (such as the maximally mixed state or diagonal states $\rho=\sum_{j_1, \cdots, j_n}p_{j_1\cdots{}j_n}|j_1\cdots{}j_n\rangle\langle j_1\cdots{}j_n|$ with the probability distribution $\{p_{j_1\cdots{}j_n}\}$) satisfying $p\rho_0+(1-p)\rho\in {\cal B}_d$ for any $\rho\in {\cal B}_d$ and $p\in (0,1)$. Without the convexity \cite{Rudin}, it is difficult to verify a general robust entanglement. The problem may be simplified for special states. One is the generalized $d$-dimensional Dicke state on Hilbert space $\otimes_{j=1}^n\mathbb{H}_{A_j}$ \cite{Toth} given by
\begin{eqnarray}
|D_{k,n}\rangle_{A_1\cdots{}A_n}=\sum_{j_1+\cdots{}+j_n=k}\alpha_{j_1\cdots{}j_n}|j_1 \cdots{}j_n\rangle,
\label{eqn2}
\end{eqnarray}
where $k$ denotes the number of excitations, $n$ denotes the number of particles, $\alpha_{j_1\cdots{}j_n}$'s satisfy $\sum_{j_1,\cdots{},j_n}\alpha_{j_1\cdots{}j_n}^2=1$ and $\alpha_{j_1\cdots{}j_n}\not=0$ for any $j_1, \cdots, j_n$.

\textbf{Result 1}. Any generalized Dicke state in Eq.(\ref{eqn2}) is robust entanglement.

\textbf{Proof}. Note that $|D_{k,n}\rangle_{A_1\cdots{}A_n}$ is a generalized symmetric state which means that after any permutation $S$ of particles the final state $S|D_{k,n}\rangle$ has the form similar to Eq.(\ref{eqn2}) (which may have different superposed amplitudes). This is weaker than the permutationally symmetry that requires the state to be invariant under any permutations of particles. From the generalized symmetry of Dicke states $|D_{k,n}\rangle$, it only needs to consider the reduced state of $\rho_{(A_{j+1}\cdots{} A_n)}={\cal E}_{A_{j+1}\cdots A_n}(|D_{k,n}\rangle\langle D_{k,n}|)$ for $j=2, \cdots, n-1$.

Take $n=3$ as an example. Note that
\begin{eqnarray}
|D_{k,3}\rangle_{A_1A_2A_3}&=& \sum_{j_1+j_2=k-j_3}\alpha_{j_1j_2}|j_1j_2\rangle_{A_1A_2}
\sum_{j_3}\alpha_{j_3}|j_3\rangle_{A_3}
\nonumber
\\
&=&\sum_{j_3}\alpha_{j_3}|j_3\rangle_{A_3}|D_{k-j_3,2}\rangle_{A_1A_2},
\label{BB1}
\end{eqnarray}
where $|D_{k-j_3,2}\rangle$ is a generalized Dicke state on Hilbert space $\mathbb{H}_{A_1}\otimes \mathbb{H}_{A_2}$ and is defined by
\begin{eqnarray}
|D_{k-j_3,2}\rangle_{A_1A_2}=\sum_{j_1+j_2=k-j_3}\alpha_{j_1j_2}|j_1j_2\rangle.
\end{eqnarray}
The density matrix of $\rho_{(A_3)}$ is given by
\begin{eqnarray}
\rho_{(A_3)}=\sum_{j_3}\alpha_{j_3}^2|D_{k-j_3,2}\rangle_{A_1A_2}\langle D_{k-j_3,2}|.
\label{BB2}
\end{eqnarray}

From Eq.(\ref{BB2}), for any $j_3$ and $j_3'$ with $j_3\not=j_3'$, both the states of $|D_{k-j_3,2}\rangle$ and $|D_{k-j_3',2}\rangle$ are generalized Dicke states with excitations $k-j_3$ and $k-j_3'$ satisfying $k-j_3\not=k-j_3'$. This implies that $|D_{k-j_3,2}\rangle$ and $|D_{k-j_3',2}\rangle$ are defined on different subspaces $\mathcal{H}_{1} \subseteq \mathbb{H}_{A_1}\otimes \mathbb{H}_{A_2}$ and $\mathcal{H}_{2}\subseteq \mathbb{H}_{A_1}\otimes \mathbb{H}_{A_2}$ respectively, where $\mathcal{H}_{1}$ is spanned by the basis states $\{|j_1j_2\rangle, j_1+j_2=k-j_3, j_1,j_2\geq0\}$, $\mathcal{H}_{2}$ is spanned by the basis states $\{|j_1j_2\rangle|j_1+j_2=k-j_3',j_1,j_2\geq0\}$, and $\mathcal{H}_{1}\cap \mathcal{H}_2=\emptyset$. Hence, $\rho_{(A_3)}$ in Eq.(\ref{BB2}) is a bipartite entanglement if there is at least one integer $j_3$ such that $|D_{k-j_3,2}\rangle$ is a bipartite entanglement from Eq.(\ref{eqn0}). This can be proved for any $d\geq 2$ ($d$ is the local dimension of each system $A_j$). In fact, for $d=2$, $|D_{1,2}\rangle=\alpha_{01}|01\rangle+\alpha_{10}|10\rangle)$ is a bipartite entanglement for any $\alpha_{01}\alpha_{10}\not=0$. For $d>2$, we get that $|D_{k-j_3,2}\rangle$ is an entangled Dicke state for any $k$ with $j_3<k\leq d$ \cite{Luo2020}. This completes the proof for $n=3$.

Similar proof holds for general integers of $n$ and $k$. In fact, from Eq.(\ref{BB1}) we have the following decomposition
\begin{eqnarray}
|D_{k,n}\rangle
&=& \sum_{\sum_{i=1}^sj_i=k-\sum_{t=s+1}^nj_{t}}\!\!\!\!\!\!\!\!\alpha_{j_1\cdots{}j_s}
|j_1\cdots{}j_s\rangle_{A_1\cdots{}A_s}
\nonumber
\\
&&\otimes\sum_{j_{s+1}, \cdots, j_n}\alpha_{j_{s+1}\cdots{}j_n}|j_{s+1}\cdots{}j_n\rangle_{A_{s+1}\cdots{}A_n}
\nonumber
\\
&=&\sum_{j_{s+1}, \cdots, j_n}\alpha_{j_{s+1}\cdots{}j_n}|j_{s+1}\cdots{}j_n\rangle_{A_{s+1}\cdots{}A_n}
\nonumber
\\
&&\otimes|D_{k-\sum_{t=s+1}^nj_{t},s}\rangle_{A_1\cdots{}A_s}
\label{BB3}
\end{eqnarray}
where $|D_{k-\sum_{t=s+1}^nj_{t},s}\rangle_{A_1\cdots{}A_s}$ are generalized Dicke states on Hilbert space $\otimes_{j=1}^{s}\mathbb{H}_{A_{j}}$ defined by
\begin{eqnarray}
|D_{k-\sum_{t=s+1}^nj_{t},s}\rangle_{A_1\cdots{}A_s}=\sum_{k-\sum_{t=s+1}^nj_{t}=\ell_s}
\!\!\!\!\!\!\!\!
\alpha_{j_1\cdots{}j_s}|j_1\cdots{}j_s\rangle
\label{DK}
\end{eqnarray}
with $\sum_{i=1}^sj_i=\ell_s$.

After $|D_{k,n}\rangle$ passing through the particle-lose channel ${\cal E}_{A_{s+1}\cdots{}A_n}(\cdot)$, the reduced state of $\rho_{(A_{s+1}\cdots{}A_n)}$ is given by
\begin{eqnarray}
&&\rho_{(A_{s+1}\cdots{}A_n)}
\nonumber
\\
\!\!\!\!\!\!
&=&{\cal E}_{A_{s+1}\cdots{}A_n}(|D_{k,n}\rangle\langle D_{k,n}|)
\nonumber
\\
\!\!\!\!\!\!&=&\!\!\!\!\!\!
\sum_{j_{s+1},\cdots, j_n}\!\!\!\!\!\!
\alpha_{j_{s+1}\cdots{}j_n}^2|D_{k-\sum_{t=s+1}^nj_{t},s}\rangle_{A_1\cdots{}A_s}\langle D_{k-\sum_{t=s+1}^nj_{t},s}|
\nonumber
\\
\label{BB4}
\end{eqnarray}

Similar to the case of $n=3$, a simple fact is as follows. For each pair of $\ell$ and $\ell'$ with $\ell\not=\ell'$, the nonnegative solutions of $i_{s+1}+\cdots +i_n=\ell$ and $j_{s+1}+\cdots +j_n=\ell'$ are different, that is,
\begin{eqnarray}
{\cal S}_{\ell}\cap{\cal S}_{\ell'}=\emptyset,
\end{eqnarray}
where ${\cal S}_{\ell}$ and ${\cal S}_{\ell'}$ are given respectively by
\begin{eqnarray}
&&{\cal S}_{\ell}=\{(i_{s+1},\cdots,i_n)|\sum_{t=s+1}^ni_{t}=\ell\},
\\
&&{\cal S}_{\ell'}=\{(j_{s+1},\cdots,j_n)|\sum_{t=s+1}^nj_t=\ell'\}.
\end{eqnarray}
This yields to $k-\ell\not=k-\ell'$. So, all the nonnegative solutions of $i_{1}+\cdots +i_s=k-\ell$ and $j_{1}+\cdots +j_s=k-\ell'$ are different, that is,
\begin{eqnarray}
{\cal S}_{k-\ell}\cap{\cal S}_{k-\ell'}=\emptyset,
\end{eqnarray}
where ${\cal S}_{\ell}$ and ${\cal S}_{\ell'}$ are given respectively by
\begin{eqnarray}
&&{\cal S}_{k-\ell}=\{(i_{},\cdots,i_s)|\sum_{t=1}^si_{t}=k-\ell\},
\\
&&{\cal S}_{k-\ell'}=\{(j_{1},\cdots,j_s)|\sum_{t=1}^sj_t=k-\ell'\}.
\end{eqnarray}
It follows $\langle i_{}\cdots{}i_s|j_{}\cdots{}j_s\rangle=0$ for $(i_{},\cdots,i_s)\in {\cal S}_{k-\ell}$ and $(j_{},\cdots,j_s)\in {\cal S}_{k-\ell'}$.
Denote $\mathcal{H}_k$ as the subspace spanned by the basis states $\{|j_1\cdots{}j_s\rangle|\sum_{t=s+1}^nj_t=\ell, j_1, \cdots, j_s\geq0\}$ and $\mathcal{H}_k'$ as the subspace spanned by the basis states $\{|j_1\cdots{}j_s\rangle|\sum_{t=1}^sj_t=\ell',j_1, \cdots, j_s\geq0\}$. We have proved $\mathcal{H}_k\cap \mathcal{H}_k'=\emptyset$ for each pair of $\ell\not=\ell'$.  This implies that for $\ell\not=\ell'$ the states of $|D_{k-\ell,s}\rangle$ and $|D_{k-\ell',s}\rangle$ in Eq.(\ref{BB4}) are generalized Dicke states defined on different subspaces $\mathcal{H}_k$ and $\mathcal{H}_k'$, respectively. Hence, $\rho_{(A_{s+1}\cdots{}A_n)}$ is a genuinely $s$-partite entanglement if and only if there is at least one integer  $\ell$ with $\ell \leq k$ such that $|D_{k-\ell,s}\rangle$ is a genuinely $s$-partite entanglement \cite{Sy}. This proves the result for $d\geq 2$ and $k\geq 1$ by using the recent result \cite{Luo2020} that states any generalized Dicke state of $|D_{k-\ell,s}\rangle$ is genuinely multipartite entangled in the biseparable model \cite{Sy}. Hence, it has proved the result.$\Box$

In what follows, consider the superposition of generalized Dicke states as
\begin{eqnarray}
|\Phi\rangle_{A_1\cdots{}A_n}=\sum_{k}\beta_{k}|D_{k,n}\rangle,
\label{BB5}
\end{eqnarray}
where $\beta_k$'s satisfies $\sum_k\beta_k^2=1$ and $|D_{k,n}\rangle$ are defined in Eq.(\ref{eqn2}). Note that a fully separable state of $|\phi\rangle^{\otimes n}$ has the decomposition in Eq.(\ref{BB5}) for any single-particle state $|\phi\rangle$, where $\beta_{k}$'s depend on the specific form of $|\phi\rangle$. This kind of states is not entangled and then not robust entanglement. Hence, in what follows, we assume that the state in Eq.(\ref{BB5}) is an $n$-partite genuinely entanglement \cite{Sy} (not any fully separable state). With this assumption, from Result 1 we can prove the state in Eq.(\ref{BB5}) is robust entanglement in the present model for any $\beta_k$'s.

\textbf{Corollary 1}. For a given state in Eq.(\ref{BB5}), it is robust entanglement if it is a genuinely $n$-partite entanglement in the biseparable model.

\textbf{Proof}. From Eq.(\ref{eqn2}), $|D_{k,n}\rangle$'s in Eq.(\ref{BB5}) are generalized Dicke states with different excitations. This means that $|D_{k,n}\rangle$'s are defined on different subspaces of $\otimes_{j=1}^n\mathbb{H}_{A_j}$. From Eqs.(\ref{BB3}) and (\ref{BB5}), it follows that
\begin{eqnarray}
\rho_{(A_{s+1}\cdots{}A_{n})}
&=&\sum_{k}\beta_k^2\sum_{j_{s+1},\cdots, j_n}
\alpha_{j_{s+1}\cdots{}j_n}^2|D_{k-\sum_{t=s+1}^nj_{t},s}\rangle
\nonumber
\\
&&\times\langle D_{k-\sum_{t=s+1}^nj_{t},s}|
\label{BB6}
\end{eqnarray}
where $\rho_{(A_{s+1}\cdots{}A_{n})}$ is the density matrix of the particles $A_1, \cdots, A_s$ after the state in Eq.(\ref{BB5}) passing through the particle-lose channel ${\cal E}_{A_{s+1}\cdots{}A_n}(\cdot)$, and $|D_{k-\sum_{t=s+1}^nj_{t},s}\rangle$ are defined in Eq.(\ref{DK}). Similar to the analysis after Eq.(\ref{BB4}), for any pair of  $k$ and $k'$ with $k\not=k'$, the states of $|D_{k-\ell,s}\rangle$ and $|D_{k'-\ell,s}\rangle$ have different excitations $k-\ell\not=k'-\ell$. This means $|D_{k-\ell,s}\rangle$ and $|D_{k'-\ell,s}\rangle$ are generalized Dicke states defined on subspaces $\mathcal{H}_k,\mathcal{H}_k'\subseteq \otimes_{j=1}^n\mathbb{H}_{A_j}$ with $\mathcal{H}_k\cap \mathcal{H}_k'=\emptyset$, where $\mathcal{H}_k$ denotes the subspace spanned by the basis states $\{|j_1\cdots{}j_s\rangle|j_1+\cdots+j_s=k-\ell, j_1, \cdots, j_s\geq0\}$ and $\mathcal{H}_k'$ denotes the subspace spanned by the basis states $\{|j_1\cdots{}j_s\rangle|j_1+\cdots{}+j_s=k-\ell',j_1, \cdots, j_s\geq0\}$. This implies that $\rho_{(A_{s+1}\cdots{}A_{n})}$ in Eq.(\ref{BB6}) is a genuinely $s$-partite entanglement if and only if one of Dicke states $|D_{k-\ell,s}\rangle$ is a genuinely $n$-partite entanglement for some integers $\ell$ and $k$ from the definition in Eq.(\ref{eqn0}). This is proved by using two facts as follows. One is from Result 1 which states all the generalized Dicke states of $|D_{k-\ell,s}\rangle$ in Eq.(\ref{DK}) are genuinely $s$-partite entangled states in the biseparable model \cite{Sy}. The other is from the assumptions in Eqs.(\ref{eqn2}) and (\ref{BB5}), that is, all the parameters of $\alpha_{j_{s+1}\cdots{}j_n}$ in Eq.(\ref{eqn2}) satisfy $\alpha_{j_{s+1}\cdots{}j_n}\not=0$; and there is at least one parameter $\beta_k$ in Eq.(\ref{BB5}) with $\beta_k\not=0$. Hence, from Eq.(\ref{BB6}), we have completed the proof. $\Box$

When $\alpha_{j_1\cdots{}j_n}$'s are all equal, $|D_{k,n}\rangle$ in Eq.(\ref{BB5}) become Dicke states \cite{Toth} which are genuinely multipartite entangled \cite{DPR,Dicke4,TDS,Sy}. Generally, $|D_{k,n}\rangle$ cannot be generated by using entangled states which have no more than $n-1$ particles assisted by CPTP mappings. This implies that they are genuinely network entangled \cite{Luo2020} and genuinely $n$-partite entangled \cite{Sy}. Result 1 and Corollary 1 provide two kinds of non-symmetric states that are robust against particle-loss.

\section{Robust entangled quantum networks}

An $n$-partite quantum network ${\cal N}_q$ consists of independent entangled states such as Affleck-Kennedy-Lieb-Tasaki (AKLT) system \cite{AKLT} with small number of particles, as shown in Fig.\ref{fig-2}. These entangled systems show great convenience for large-scale quantum tasks \cite{Kim} with short-range experimental settings. The independence assumption of ${\cal N}_q$ is the key to activate new non-localities depending on network configurations \cite{BGP,SBP,RBB,Chav,Luo2018,Luo2019,CPV,Fri,RBBB}. Nevertheless, it may rule out specific scenarios such as cyclic networks \cite{RBBB,Luo2018}. Our goal is to characterize general quantum networks under local unitary operations or generalized CPTP mappings. This allows for regarding all the particles shared by one party as one combined particle in large Hilbert space. In this case, the particle-lose in Definition 1 means network nodes or parties in applications may be lost or offline.

\subsection{$k$-independent quantum networks}

\begin{figure}
\begin{center}
\resizebox{240pt}{280pt}{\includegraphics{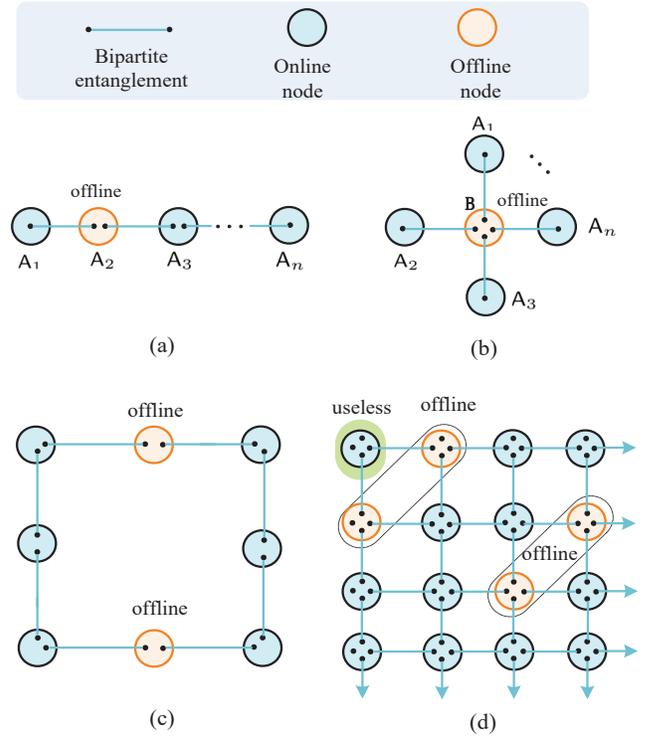}}
\end{center}
\caption{\small (Color online) Schematic $k$-independent quantum networks consisting of bipartite entangled states. (a) Chain-type network. (b) Star-type network. (c) Cyclic network. (d) Planer network used in measurement-based quantum computation \cite{Cluster}. These networks are particle-lose separable.}
\label{fig-2}
\end{figure}

Consider a general network ${\cal N}_q$ consisting of parties $\textsf{A}_1, \cdots, \textsf{A}_n$, where each party may share some entangled states with other. ${\cal N}_q$ is a $k$-independent quantum network if there are $k$ number of parties $\textsf{A}_1, \cdots, \textsf{A}_k$ (for example) any pair of them have not pre-shared entangled states. Although this kind of quantum networks show nonlocality according to the specific Bell-type inequalities \cite{Luo2018,Luo2019}, we can prove they are not robust against node-loss. These include the chain-type network \cite{ZZH}, the star-type network and general networks \cite{Luo2018}, as shown in Fig.\ref{fig-2}.

\textbf{Result 2}. Any $k$-independent quantum network ${\cal N}_q$ with $k\geq2$ is particle-lose separable.

\textbf{Proof}. Consider an $n$-partite $k$-independent quantum network ${\cal N}_q$ consisting of $n$ parties $\textsf{A}_1, \cdots, \textsf{A}_n$, where $\textsf{A}_{1},\cdots, \textsf{A}_{k}$ (for example) are independent, that is, they have no prior-shared entanglement \cite{Luo2018}. Suppose that a $k$-independent quantum network ${\cal N}_q$ consists of entangled pure states $|\Phi_1\rangle, \cdots, |\Phi_m\rangle$, where each party $\textsf{A}_j$ may share some particles in $|\Phi_l\rangle$'s with other parties. The total state of ${\cal N}_q$ is given by
\begin{eqnarray}
|\Omega\rangle_{\textsf{A}_1\cdots\textsf{A}_n}=\otimes_{j=1}^m|\Phi_j\rangle,
\label{DD1}
\end{eqnarray}
where $\textsf{A}_{1},\cdots, \textsf{A}_{k}$ may share some entangled states with other parties $\textsf{A}_{k+1}, \cdots, \textsf{A}_{n}$, who can share some entangled states with each other.

After losing all the particles shared by the parties $\textsf{A}_{k+1}, \cdots, \textsf{A}_{n}$, the joint state shared by $\textsf{A}_{1},\cdots, \textsf{A}_{k}$ is given by
\begin{eqnarray}
\rho_{(\textsf{A}_{k+1}\cdots \textsf{A}_n)}={\cal E}_{\textsf{A}_{k+1}\cdots{}\textsf{A}_n}(|\Omega\rangle\langle \Omega|),
\label{DD2}
\end{eqnarray}
where ${\cal E}_{\textsf{A}_{k+1}\cdots{}\textsf{A}_n}(\cdot)$ denotes the particle-lose channel associated with all the particles owned by $\textsf{A}_{k+1}, \cdots{}, \textsf{A}_n$.

A simple fact is that the local operations of all the parties $\textsf{A}_{k+1}, \cdots, \textsf{A}_n$ do not change the joint state of $\textsf{A}_{1},\cdots, \textsf{A}_{k}$, that is,
\begin{eqnarray}
\rho_{(\textsf{A}_{k+1}\cdots \textsf{A}_n)}&=&
{\cal E}_{\textsf{A}_{k+1}\cdots{}\textsf{A}_n}( \mathbbm{1}_{\textsf{A}_{1}\cdots\textsf{A}_{k}}\otimes(\otimes_{j=k+1}^nU_j)
|\Omega\rangle\langle \Omega|
 \nonumber
 \\
 &&\times \mathbbm{1}_{\textsf{A}_{1}\cdots\textsf{A}_{k}}\otimes(\otimes_{j=k+1}^nU^\dag_j))
\label{DD3}
\end{eqnarray}
where $\mathbbm{1}_{\textsf{A}_{1}\cdots\textsf{A}_{k}}$ denotes the identity operator on the joint system of $\textsf{A}_{1},\cdots, \textsf{A}_{k}$, and $U_j$ denotes the local unitary operation performed by the party $\textsf{A}_j$, $j\in \{k+1, \cdots, n\}$. From Eq.(\ref{DD3}), it follows that $\rho_{(\textsf{A}_{k+1}\cdots \textsf{A}_n)}$ is fully separable because $\textsf{A}_{k+1},\cdots, \textsf{A}_{n}$ do not share any entanglement from the assumption of $k$-independence. This implies that $|\Omega\rangle$ in Eq.(\ref{DD1}) is robust entangled in the present model.

Similar proof holds for any $k$-independent quantum network consisting of general entangled states including mixed entangled states. This completes the proof. $\Box$

\textit{Example 1}. Consider an $n$-partite chain-type network ${\cal N}_q$ shown in Fig.\ref{fig-2}(a), where each pair of two adjacent parties $\textsf{A}_{j}$ and $\textsf{A}_{j+1}$ shares one EPR state $|\Phi_j\rangle=\frac{1}{\sqrt{2}}(|00\rangle+|11\rangle)$. The total state of ${\cal N}_q$ is given by $|\Omega\rangle_{\textsf{A}_1\cdots\textsf{A}_n}=\otimes_{j=1}^{n-1}|\Phi_j\rangle$ which can be regarded as a pure state on Hilbert space $\otimes_{j=1}^n\mathbbm{H}_{\textsf{A}_j}$, where $\mathbbm{H}_{\textsf{A}_1}$ and $\mathbbm{H}_{\textsf{A}_n}$ are 2-dimensional spaces while $\mathbbm{H}_{\textsf{A}_j}$ with $j\in \{2, \cdots, n-1\}$ are 4-dimensional spaces. Take $n=3$ as example, the total state is given by
\begin{eqnarray}
|\Omega\rangle_{\textsf{A}_1\textsf{A}_2\textsf{A}_3}&=&\frac{1}{2\sqrt{2}}
(|0\rangle_{\textsf{A}_1}|00\rangle_{\textsf{A}_2}|0\rangle_{\textsf{A}_3}
+|0\rangle_{\textsf{A}_1}|01\rangle_{\textsf{A}_2}|1\rangle_{\textsf{A}_3}
 \nonumber
 \\
 &&+|1\rangle_{\textsf{A}_1}|10\rangle_{\textsf{A}_2}|0\rangle_{\textsf{A}_3}
+|1\rangle_{\textsf{A}_1}|11\rangle_{\textsf{A}_2}|1\rangle_{\textsf{A}_3})
\label{DD6}
\end{eqnarray}
Here, $\textsf{A}_1$ and $\textsf{A}_3$ are independent parties. After passing through the particle-lose channel ${\cal E}_{\textsf{A}_{2}}(\cdot{})$, the reduced state of $\textsf{A}_1$ and $\textsf{A}_3$ is given by
\begin{eqnarray}
\rho_{(\textsf{A}_{2})}=\frac{1}{4}\sum_{i,j=0}^1|ij\rangle\langle{}ij|)
\label{DD7}
\end{eqnarray}
which is the maximally mixed state. Similarly, for general case of $n\geq 3$, all the parties $\textsf{A}_j$'s with even $j$ (or odd $j$) are independent. The reduced state after ${\cal N}_q$ passing through all the particle-lose channels ${\cal E}_{\textsf{A}_{j}}(\cdot{})$ with even $j$'s is fully separable.

\textit{Example 2}. Consider an $n+1$-partite star-type network ${\cal N}_q$ shown in Fig.\ref{fig-2}(b), where each outer party $\textsf{A}_{j}$ shares one EPR state $|\Phi\rangle=\frac{1}{\sqrt{2}}(|00\rangle+|11\rangle)$ with the center party $\textsf{B}$. The total state of ${\cal N}_q$ is given by
\begin{eqnarray}
|\Omega\rangle_{\textsf{A}_1\cdots\textsf{A}_n\textsf{B}}=\frac{1}{\sqrt{2^n}}\!\!\!\!
\sum_{j_1,\cdots, j_n=0}^1\!\!\!\!|j_1\cdots{}j_n\rangle_{\textsf{A}_1\cdots\textsf{A}_n}
|j_1\cdots{}j_n\rangle_{\textsf{B}}
\label{DD7}
\end{eqnarray}
which can be regarded as a pure state on Hilbert space $\mathbbm{H}_{B}\otimes(\otimes_{j=1}^n\mathbbm{H}_{\textsf{A}_j})$, where $\mathbbm{H}_{\textsf{A}_j}$ are 2-dimensional spaces while $\mathbbm{H}_{\textsf{B}}$ is $2^n$-dimensional space. Here, all the parties of $\textsf{A}_j$'s are independent parties. After passing through the particle-lose channel ${\cal E}_{\textsf{B}}(\cdot{})$, the reduced state of $\textsf{A}_1, \cdots, \textsf{A}_n$ is given by
\begin{eqnarray}
\rho_{(\textsf{B})}=\frac{1}{2^n}\sum_{j_1,\cdots, j_n=0}^1|j_1\cdots{}j_n\rangle_{\textsf{A}_1\cdots\textsf{A}_n}\langle j_1\cdots{}j_n|
\label{DD8}
\end{eqnarray}
which is the maximally mixed state.

Another example is cyclic network in Fig.\ref{fig-2}(c) or planer networks Fig.\ref{fig-2}(d). Result 2 shows new feature of all the $k$-independent quantum networks, which depends only on the independence assumption of the parties, but not the shared entangled states. The lack of robustness against particle-loss may rule out some specific applications such as measurement-based quantum computation \cite{Cluster} using $k$-independent quantum networks.

\begin{figure}
\begin{center}
\resizebox{230pt}{160pt}{\includegraphics{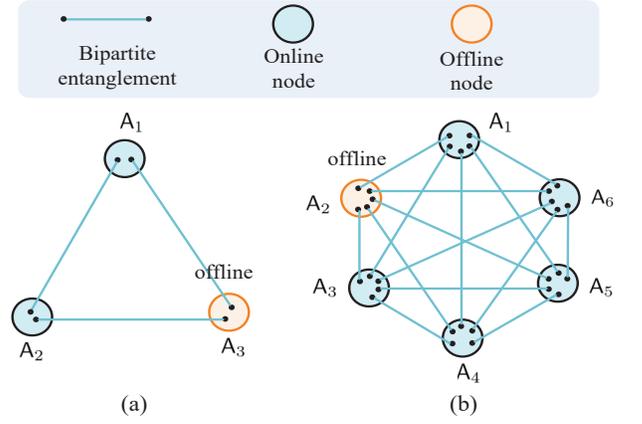}}
\end{center}
\caption{\small (Color online) Schematic completely connected quantum networks consisting of bipartite entangled states. (a) Triangle network. (b) Completed network with six parties. }
\label{fig-4}
\end{figure}

\subsection{Completely connected quantum networks}

Different from these scenarios in Result 2 there are other networks which are robust against particle-lose, as shown in Fig.\ref{fig-4}.

\textbf{Definition 2}. An $n$-partite quantum network ${\cal N}_q$ is a \textit{completely connected} if each pair of two parties shares at least one entanglement.

Suppose that ${\cal N}_q$ consists of bipartite entangled pure states \cite{EPR,Gis} given by
\begin{eqnarray}
|\phi\rangle_{AB}=\sum_{j}\beta_j|jj\rangle
\label{CC}
\end{eqnarray}
on a finite-dimensional Hilbert space $\mathbbm{H}_A\otimes\mathbbm{H}_B$, and generalized Dicke states in Eq.(\ref{eqn2}), where $\beta_j$'s satisfy $\sum_{j}\beta_j^2=1$ and $\beta_j\not=0$ for at least two integers $j$. We have the following result.

\textbf{Result 3}. Any completely connected quantum network is robust entanglement.

\textbf{Proof}. We firstly prove that the total state of any completely connected quantum network ${\cal N}_q$ is an $n$-partite genuinely entanglement in the biseparable model \cite{Sy} if ${\cal N}_q$ consists of bipartite entangled pure states and generalized Dicke states. This can be proved by using the following fact.

\textbf{Fact 1}. An $m$-partite quantum network is genuinely $m$-partite entanglement if it consists of bipartite entangled pure states.

Fact 1 can be proved by using a recent method assisted by local operations and classical communication \cite{ZDBS}, that is, an $n$-partite state is genuinely entangled if each pair of two parties can share one bipartite pure entanglement with the help of others' local measurements and classical communication.

For any $m$-particle Dicke state $|D_{k,m}\rangle$ defined similar to  Eq.(\ref{eqn2}), it is genuinely multipartite entanglement in the biseparable model \cite{Sy} from Result 1. Moreover, the final joint state of any two particles after $|D_{k,m}\rangle$ being measured using proper projections on other particles is a bipartite entanglement \cite{ZDBS}. Hence, from the generalized entanglement swapping \cite{ZZH}, each pair of two parties $\textsf{A}_i$ and $\textsf{A}_j$ in ${\cal N}_q$ can share one bipartite entanglement with the help of others' local measurement and classical communication. This means the total state of ${\cal N}_q$ is a genuinely $n$-partite entanglement. The following proof is completed by two cases.

\textbf{Case 1}. An $n$-partite completely connected quantum network consists of bipartite entangled pure states.

In this case, ${\cal N}_q\cup\{|\phi\rangle\}$ is robust entanglement if ${\cal N}_q$ is robust entanglement, where ${\cal N}_q\cup\{|\phi\rangle\}$ denotes a new network by adding one bipartite entanglement $|\phi\rangle$ shared by one party or two parties. With this fact, it only needs to consider the simplest case of the completely connected network ${\cal N}_q$ on which each pair of $\textsf{A}_i$ and $\textsf{A}_j$ shares one bipartite entanglement $|\phi_{i;j}\rangle$ defined by
\begin{eqnarray}
|\phi_{i;j}\rangle=\sum_{k}\alpha_{k;ij}|kk\rangle,
\label{C4}
\end{eqnarray}
where the local dimension of each system satisfies $d\geq 2$, $\alpha_{k;ij}$'s satisfy $\sum_k\alpha_{k;ij}^2=1$ and $\alpha_{k;ij}\not=0$ for at least two different integers $k$. The total system of ${\cal N}_q$ is given by
\begin{eqnarray}
|\Phi\rangle_{\textsf{A}_{1}\cdots\textsf{A}_{n}}=\otimes_{i,j,i\not=j}|\phi_{i; j}\rangle
\label{C5}
\end{eqnarray}

Consider a subset $S=\{\textsf{A}_{i_1}, \cdots, \textsf{A}_{i_k}\}\subset\{\textsf{A}_{1}, \cdots, \textsf{A}_n\}$ and $\overline{S}$ being the complement set of $S$. The joint state after ${\cal N}_q$ passing through the particle-lose channel ${\cal E}_{\overline{S}}(\cdot):={\cal E}_{\textsf{A}_i\in \overline{S}}(\cdot)$ is given by
\begin{eqnarray}
 \rho_{(\overline{S})}=(\otimes_{s\not=t, s,t\in \{i_1,\cdots, i_k\}}|\phi_{s;t}\rangle\langle \phi_{s;t}|)\otimes_{j=1}^k\rho_{i_j},
\label{Cc6}
\end{eqnarray}
where $\rho_{i_j}$ denotes the local state shared by $\textsf{A}_{i_j}$ after losing all the particles entangled with $\textsf{A}_{i_j}$, which is given by $\rho_{i_j}=\otimes_{\tau\not\in\{i_1,\cdots,i_k\}}{\rm Tr}_{\textsf{A}_{i_\tau}}(|\phi_{i_j; \tau}\rangle\langle \phi_{i_j;\tau}|)$, and ${\rm Tr}_{\textsf{A}_i}(\cdot)$ denotes the trace operator by tracing out all the particles shared by $\textsf{A}_i$. From Eq.(\ref{Cc6}) it only needs to consider $\otimes_{s\not=t, s,t\in \{i_1,\cdots, i_k\}}|\phi_{s;t}\rangle\langle \phi_{s;t}|$ while $\otimes_{j=1}^k\rho_{i_j}$ is fully separable.

In fact, for any $S$ and $S'$ with $S\not=S'$ and $|S|=|S'|$, the density matrices of $\rho_{(\overline{S})}$ and $\rho_{(\overline{S'})}$ have decompositions similar to Eq.(\ref{Cc6}). This implies the generalized symmetry of $|\Phi\rangle$ in Eq.(\ref{C5}). Hence, it only needs to consider $S=\{\textsf{A}_1, \cdots, \textsf{A}_k\}$ for $k=2, \cdots, n-1$. Note that the subnetwork consisting of $\textsf{A}_{1}, \cdots, \textsf{A}_{k}$ is connected from Definition 2 after losing all the particles owned by $\textsf{A}_{k+1}, \cdots, \textsf{A}_{n}$. This implies a chain-type subnetwork consisting of $\textsf{A}_{1}, \cdots, \textsf{A}_{k}$. From Fact 1, $\rho_{(\overline{S})}$ in Eq.(\ref{Cc6}) is a genuinely $k$-partite entanglement \cite{ZDBS}. It means that the total state of ${\cal N}_q$ is robust entanglement.

\textbf{Case 2}. An $n$-partite completely connected quantum network consists of bipartite entangled pure states and generalized Dicke states.

Similar to Case 1, it only needs to consider the simplest case that each pair of $\textsf{A}_i$ and $\textsf{A}_j$ shares one bipartite entanglement $|\phi_{ij}\rangle$ defined in Eq.(\ref{C4}) or generalized Dicke state $|D_{k_{ij},m_{ij}}\rangle$ in Eq.(\ref{eqn2}). The total system of ${\cal N}_q$ is given by
\begin{eqnarray}
|\Psi\rangle_{\textsf{A}_{1}\cdots\textsf{A}_{n}}=
\otimes_{i,j}|\phi_{i;j}\rangle\otimes_{s,t}|D_{k_{st},m_{st}}\rangle
\label{C7}
\end{eqnarray}

For a given set $S=\{\textsf{A}_{j_1}, \cdots, \textsf{A}_{j_k}\}\subset\{\textsf{A}_{1}, \cdots, \textsf{A}_n\}$ and its complement set $\overline{S}$, the joint state after  ${\cal N}_q$ passing through the particle-lose channel ${\cal E}_{\overline{S}}(\cdot)={\cal E}_{\textsf{A}_i\in \overline{S}}(\cdot)$ is given by
\begin{eqnarray}
\rho_{(\overline{S})}&=&{\cal E}_{\overline{S}}(|\Psi\rangle\langle\Psi|)
\nonumber
\\
&=&\otimes_{s\not=t, s,t\in \{j_1,\cdots, j_k\}}|\phi_{s;t}\rangle\langle \phi_{s;t}|
\nonumber\\
&& \otimes(\otimes_{u\not=v, u,v\in \{j_1,\cdots, j_k\}} \rho_{u,v})\otimes(\otimes_{\ell=1}^k\varrho_{(j_\ell)}),
\label{C8}
\end{eqnarray}
where $\rho_{u,v}$ as bipartite states shared by the parties $\textsf{A}_s$ and $\textsf{A}_t$ are the remained states of generalized Dicke states after losing all the particles being not shared by $\textsf{A}_s$ and $\textsf{A}_t$  and $\varrho_{j_\ell}$ denotes the local state shared by $\textsf{A}_{j_\ell}$ after losing all the particles entangled with $\textsf{A}_{j_\ell}$. From Eq.(\ref{BB3}) it is sufficient to consider $(\otimes_{s\not=t, s,t\in \{j_1,\cdots, j_k\}}|\phi_{s;t}\rangle\langle \phi_{s;t}|)\otimes(\otimes_{u\not=v, u,v\in \{j_1,\cdots, j_k\}} \rho_{u,v})$ while $\otimes_{\ell=1}^k\varrho_{j_\ell}$ is fully separable.

From Result 1, any generalized Dicke state is robust entanglement. After passing through a complete positive preserve trace channel ${\cal E}_{\overline{S}}(\cdot)$, ${\cal N}_q$ is connected, where each pair of $\textsf{A}_s, \textsf{A}_t\in S$ shares one bipartite entangled pure state $|\phi_{s;t}\rangle$ or the remained entanglement $\rho_{s,t}$ from generalized Dicke states. From Fact 1, $\rho_{\overline{S}}$ is a genuinely $k$-partite entanglement \cite{ZDBS}. It means that the total state of ${\cal N}_q$ is robust entanglement in the present model. This has completed the proof. $\Box$

Similar result may be proved for any completely connected quantum network consisting of other pure entangled states or mixed entangled states \cite{ZDBS}.

\textit{Example 3.}  Consider an $3$-partite completely connected quantum network ${\cal N}_q$ \cite{EPR,RBBB} shown in Fig.\ref{fig-4}(a), where each pair shares one EPR state $|\phi\rangle=\frac{1}{\sqrt{2}}(|00\rangle+|11\rangle)$. The total state of ${\cal N}_q$ is given by
\begin{eqnarray}
|\Omega\rangle_{\textsf{A}_1\textsf{A}_2\textsf{A}_3}&=&\frac{1}{2\sqrt{2}}
(|00\rangle_{\textsf{A}_1}|00\rangle_{\textsf{A}_2}|00\rangle_{\textsf{A}_3}
\nonumber\\
&&+|10\rangle_{\textsf{A}_1}|00\rangle_{\textsf{A}_2}|01\rangle_{\textsf{A}_3}
\nonumber\\
&&+|00\rangle_{\textsf{A}_1}|01\rangle_{\textsf{A}_2}|10\rangle_{\textsf{A}_3}
\nonumber\\
&&+|10\rangle_{\textsf{A}_1}|01\rangle_{\textsf{A}_2}|11\rangle_{\textsf{A}_3}
\nonumber\\
&&+|01\rangle_{\textsf{A}_1}|10\rangle_{\textsf{A}_2}|00\rangle_{\textsf{A}_3}
\nonumber\\
&&+|11\rangle_{\textsf{A}_1}|10\rangle_{\textsf{A}_2}|01\rangle_{\textsf{A}_3}
\nonumber\\
&&+|01\rangle_{\textsf{A}_1}|11\rangle_{\textsf{A}_2}|10\rangle_{\textsf{A}_3}
\nonumber\\
&&+|11\rangle_{\textsf{A}_1}|11\rangle_{\textsf{A}_2}|11\rangle_{\textsf{A}_3}
)
\label{DD7}
\end{eqnarray}
which can be regarded as a pure state on Hilbert space $\otimes_{j=1}^3\mathbbm{H}_{A_j}$, where $\mathbbm{H}_{A_j}$ are 4-dimensional spaces. Here, all the parties of $\textsf{A}_j$'s are independent parties. The reduced state $\rho_{(\textsf{A}_1)}$ is given by
\begin{eqnarray}
\rho_{(\textsf{A}_1)}=
\frac{1}{4}\mathbbm{1}_{\textsf{A}_2}\otimes
|\phi\rangle_{\textsf{A}_2\textsf{A}_3}\langle\phi|\otimes \mathbbm{1}_{\textsf{A}_3}
\label{DD8}
\end{eqnarray}
which is an entanglement for two parties $\textsf{A}_2$ and $\textsf{A}_3$, where
$\mathbbm{1}_{\textsf{A}_i}$ denotes an identity operator. Similar proofs hold for other reduced states $\rho_{(\textsf{A}_2)}$ and $\rho_{(\textsf{A}_3)}$. Hence, the total state of ${\cal N}_q$ is robust entanglement in the present model. Moreover, this can be extended to other quantum network shown in Fig.\ref{fig-4}(b).

Result 3 shows new insight on quantum networks going beyond the $k$-independent networks \cite{Luo2018}. Note that ${\cal N}_q$ is separable in the recent model \cite{NWR,Kraft,Luo2020} if each Dicke state is at most $n-1$-partite entangled. This presents another feature of completely connected quantum networks \cite{Luo2020}.

\section{Strong nonlocality of robust multipartite entanglement}

The present model is useful for witnessing robust entanglement such as generalized Dicke states \cite{Toth} or completed quantum networks. A natural problem is how to verify its nonlocality from Bell experiments, similar to single entanglement of EPR, GHZ or Dicke states \cite{CHSH,GHZ,Sy} (without local tensor decompositions or a rigid definition of network local states \cite{NWR,Kraft,Luo2020}) or entangled networks \cite{RBBB,Luo2018,Luo2019} as shown in Figs.\ref{fig-2} and \ref{fig-4}. Compared with the biseparable model \cite{Sy}, one may expect stronger nonlocality in the present model. It is difficult to verify the strong nonlocality of robust entanglement due to the non-convexity of ${\cal B}_d$ and that the output state depends on the particle-loss channel ${\cal E}_S(\cdot)$ associated with specific set $S$. Our goal here is to address this problem by using new Bell-like inequalities. The main idea is inspired by the following set of Bell inequalities \cite{Sy},
\begin{eqnarray}
\left\{
\begin{array}{ll}
{\cal L}_n(\rho_{A_1,\cdots A_n})\leq c_n,
\\
{\cal L}_{k}({\cal E}_{S}(\rho))\leq c_{k}, \forall  |S|\leq k,
\end{array}
\right.
\label{eqn3}
\end{eqnarray}
where ${\cal L}_m(\cdot)$ is an $m$-partite Bell operator for verifying the genuinely $m$-partite nonlocality \cite{Sy}, $c_n$ and $c_{k}$ are constants. It is necessary to violate all the inequalities in order to verify the strong nonlocality of a robust entanglement.

\textit{Example 4}. Consider a triangle network consisting three EPR states $|\phi\rangle=\cos\theta|00\rangle+\sin\theta|11\rangle$ with $\theta\in (0,\frac{\pi}{4}]$, as shown in Fig.\ref{fig-4}(a). The nonlocality can be verified by using recent method \cite{LuoDV} with multiple measurement settings. The final state after passing through particle-lose channel ${\cal E}_{\textsf{A}_i}(\cdot)$ can be verified by violating the CHSH inequality \cite{CHSH,Loub} for any $\theta\in (0,\frac{\pi}{4}]$.

\begin{figure}
\begin{center}
\resizebox{220pt}{170pt}{\includegraphics{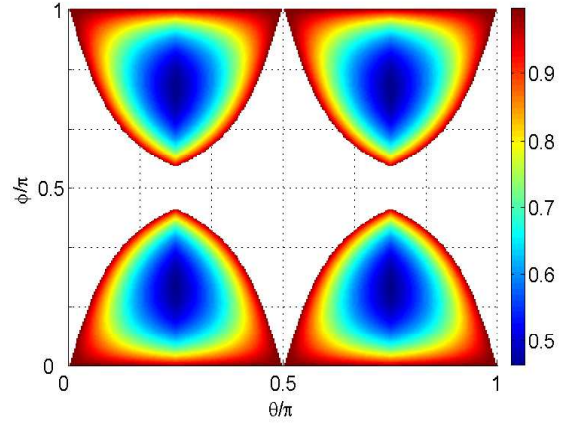}}
\end{center}
\caption{\small (Color online) The violations of the inequality (\ref{eqn5}) by using Pauil observable. Here, $\alpha=\sin\theta\cos\phi,\beta=\sin\theta\sin\phi$ and $\gamma=\cos\theta$.}
\label{fig-a3}
\end{figure}

\textit{Example 5}. Consider a generalized W state \cite{DVC} as
\begin{eqnarray}
|W\rangle_{A_1A_2A_3}=\alpha|001\rangle+\beta|010\rangle+\gamma|100\rangle
\label{eqn4}
\end{eqnarray}
with $\alpha^2+\beta^2+\gamma^2=1$. Its genuinely tripartite nonlocality can be verified by using the Svetlichny inequality \cite{Sy,AR}. Surprisingly, all the linear inequalities such as the CHSH inequality \cite{CHSH} with dichotomic settings are useless for verifying all the output states of $\rho_{(A_i)}$'s \cite{Fei,Cheng,Fine}, see Appendix E. Here, we construct a new nonlinear inequality as follows (see Appendix F),
\begin{eqnarray}
&&(\langle X_1X_2\rangle-\langle Y_1Y_2\rangle)^2+(\langle X_1Y_2\rangle+\langle Y_1X_2\rangle)^2
\nonumber\\
&&-(1-\langle{}Z_1Z_2\rangle)^2\leq 0,
\label{eqn5}
\end{eqnarray}
which holds for any separable qubit states, where $X_j$, $Y_j$ and $Z_j$ are Pauli observables on the $j$-th qubit, $j=1,2$. The maximal quantum bound is $4$. The inequality (\ref{eqn5}) provides a nonlinear entanglement witness for verifying the non-locality of $\rho_{(A_j)}$, $j=1, 2, 3$, as shown in Fig. \ref{fig-a3}. It can be further regarded as a Hardy-type inequality \cite{Hardy}. The new inequality is applicable for verifying the strong nonlocality of the generalized W states.

\begin{figure}
\begin{center}
\resizebox{240pt}{190pt}{\includegraphics{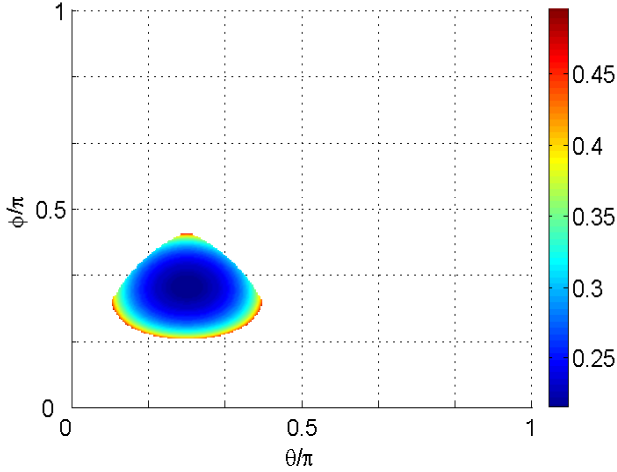}}
\end{center}
\caption{\small (Color online) The visibilities of $v$. Svetlichny inequality \cite{Sy,AR} is used for verifying noisy state $\rho$ while the inequality (\ref{eqn5}) is used for the reduced states $\rho_{(A_i)}$. Here, $\alpha=\sin\theta\cos\phi,\beta=\sin\theta\sin\phi$ and $\gamma=\cos\theta$.}
\label{fig-41}
\end{figure}

For the noisy state \cite{Werner} of
\begin{eqnarray}
\rho=v|W\rangle\langle W|+\frac{1-v}{8}\mathbbm{1}_8,
\label{a5}
\end{eqnarray}
the Svetlichny inequality \cite{Sy,AR} detects the non-locality of $\rho$ for $v\geq \min\{\frac{4}{\max_{\theta}\{2\Delta(\sin3\theta+\sin\theta)
-\sin3\theta+3\sin\theta\}},1\}$ with $\Delta=\alpha\beta+\alpha\gamma+\beta\gamma$, where $\mathbbm{1}_8$ denotes the identity operator on three qubits and $v\in (0,1)$. Fig.~\ref{fig-41} shows the visibilities of $v$ for verifying the strong nonlocalities using the Svetlichny inequality \cite{Sy,AR} and the inequality (\ref{eqn5}). The inequality (\ref{eqn5}) can be further extended for multipartite scenarios as
\begin{eqnarray}
&&4\langle(iX+Y)^{\otimes n}\rangle\langle(iX-Y)^{\otimes n}\rangle
 -(2^n-\langle (\mathbbm{1}+Z)^{\otimes n}\rangle
\nonumber
\\&&-\langle (\mathbbm{1}-Z)^{\otimes n}\rangle)^2\leq 0,
\label{multinon}
\end{eqnarray}
which holds for any biseparable qubit state \cite{Sy}. The present inequality (\ref{multinon}) is useful for verifying some entangled states \cite{Ghose09,AR} which cannot be verified by using the Svetlichny inequality \cite{Sy} under the assumption of each particle is qubit state in local hidden state model. Hence, the assumption of the present method for the nonlocality is stronger than Bell nonlocality. The proof is shown in Appendix G.

\section{Robustness-depth of multipartite entanglement}

Let ${\cal B}_b$ consist of all biseparable states \cite{Sy}. It is easy to show that ${\cal B}_{b}\subset {\cal B}_d$ on the same Hilbert space $\otimes_{j=1}^n\mathbbm{H}_{A_j}$, where ${\cal B}_d$ is defined at the beginning of Sec.III. Denote ${\cal B}_n$ as the set consisting of all the $n$-partite network local states \cite{NWR,Kraft,Luo2020} that can be generated by using $m$-partite entangled states with $m<n$, shared randomness and local operations. For each $n$-partite state $\rho\in {\cal B}_b$, it is biseparable \cite{Sy}. It implies that $\rho\in {\cal B}_d$ from Definition 1 with $|S|=n$. We get that ${\cal B}_b\subseteq{\cal B}_{d}\cap {\cal B}_{n}$. Note there are states (Examples 1 and 2) that are genuinely multipartite entangled states but not network entanglement or robust entanglement. This means ${\cal B}_b\subset{\cal B}_{d}\cap {\cal B}_{n}$. This inspires one natural problem to verify entangled states in terms of the robustness-depth.

\textbf{Definition 3}. An $n$-partite state $\rho$ has robustness-depth $k$ if ${\cal E}_{S}(\rho)$ is entangled in the biseparable model for any $S$ with $|S|\leq k-1$.

Let ${\cal B}_{d}^{(k)}$ be the set consisting of all the states with the robustness-depth less than $k$. ${\cal B}_{d}^{(k)}$ is star-convex with one center subset consisting of fully separable states, see Appendix H. For any $n$-partite state $\rho\not\in {\cal B}_{d}^{(k)}$, $\rho$ has the robustness-depth no less than $k$, that is, $\rho$ is robust entanglement against the loss of no more than $k-1$ particles. This model is useful for witnessing the robustness depth in terms of the particle loss.

\textbf{Definition 4}. An $n$-partite quantum network is $k$-connected if each party shares at most $k$ bipartite entangled states with other parties.

\textbf{Proposition 3}. The robustness-depth of any $n$-partite $k$-connected quantum network is at most $k-1$.

\textbf{Proof}. It is sufficient to show that there is one subset $S=\{\textsf{A}_{j_1}, \cdots, \textsf{A}_{j_k}\}$ such that the reduced state of $\rho_{(S)}={\cal{}E}_S(\rho)$ is biseparable, where $\rho$ denotes the total state of ${\cal N}_q$.

The main idea is from the $k$-connectedness of ${\cal N}_q$ in Definition 4. In detail, consider any party $\textsf{A}_{j}$ who shares at most $k$ bipartite entangled states with the parties $\textsf{A}_{j_1}, \cdots, \textsf{A}_{j_k}$. Now, define the subset $S=\{\textsf{A}_{j_1}, \cdots, \textsf{A}_{j_k}\}$. Consider the final state $\rho_{(S)}={\cal{}E}_{S}(\rho)$ after passing through the particle-loss channel ${\cal{} E}_{S}(\cdot{})$, which is associated with the particles owned by $\textsf{A}_{j_1}, \cdots, \textsf{A}_{j_k}$. From Definition 4, there are at most $k$ bipartite entangled states shared by $\textsf{A}_{j}$. After ${\cal N}_q$ passing through the particle-loss channel ${\cal{}E}_{S}(\cdot{})$, $\textsf{A}_{j}$ does not share entangled states with others, where each party $\textsf{A}_{j_s}$ in $S$ shares at most one entanglement with $\textsf{A}_{j}$. Here, the particle-loss channel ${\cal{}E}_{S}(\cdot{})$ is to cut all the entangled states shared with the party $\textsf{A}_{j}$. So, we get that
\begin{eqnarray}
\rho_{(S)}=\rho_{\textsf{A}_j}\otimes \varrho_{(S)|\textsf{A}_j},
\label{G1}
\end{eqnarray}
where $\rho_{\textsf{A}_j}={\cal{} E}_{B_{t_1}\cdots{}B_{t_k}}(\otimes_{s=1}^k|\Phi_{s}\rangle_{A_{t_s}B_{t_s}}\langle \Phi_s|)$ denotes the state owned by $\textsf{A}_j$, and $\varrho_{(S)|\textsf{A}_j}$ denotes the total state (after the particle-lose channel ${\cal E}_{\textsf{A}_{j_s}\not\in S}(\cdot{})$ being performed) shared by all the parties except for $\textsf{A}_j$ and all parties not in $S$. Since $\rho_{(S)}$ is separable, from Definition 4, the robustness-depth of $\rho$ is at most $k-1$. This completes the proof. $\Box$

For an $n$-partite $k$-connected quantum network ${\cal N}_q$, it is genuinely $n$-partite entangled \cite{Sy,ZDBS}, where each pair can recover one bipartite entanglement assisted by other parties' local operations and classical communication. These networks are inherent nonlocal because of the connectedness \cite{CPV}. Nevertheless, ${\cal N}_q$ is not robust against the loss of at most $k$ particles. This provides interesting examples included in ${\cal B}_{d}^{(k)}$ and yields to a general hierarchy of multipartite states in terms of the robustness-depth as
\begin{eqnarray}
{\cal B}_s\subset{\cal B}_b\subset{\cal B}_{d}^{(1)} \subset {\cal B}_{d}^{(2)}
\subset \cdots \subset {\cal B}_{d}^{(n-1)},
\label{equiv}
\end{eqnarray}
where ${\cal B}_s$ consists of all fully separable states. Due to the non-convexity of ${\cal B}_{d}^{(k)}$, it is difficult to verify the robustness-depth for general state $\rho\not\in {\cal B}_{d}^{(k)}$. Interestingly, quantum networks provide an easy example. Suppose that ${\cal N}_q$ is an $n$-partite network consisting of bipartite entangled states. The robustness-depth of ${\cal N}_q$ is then determined by its minimum degree.

\textbf{Definition 5}. For a given quantum network ${\cal N}_q$ consisting of all bipartite entangled states, the degree $\deg(\textsf{A}_j)$ of one party $\textsf{A}_j$ is the number of parties $\textsf{A}_k$ with $k\not=j$ such that $\textsf{A}_j$ and $\textsf{A}_k$ share at least one entanglement. The minimum degree of ${\cal N}_q$ is the minimal degrees of all parties, that is, $\deg({\cal N}_q)=\min\{\deg(\textsf{A}_j), \forall j\}$.

\textbf{Result 4}. The robustness-depth of ${\cal N}_q$ is given by $d=\deg({\cal N}_q)-1$, where $\deg({\cal N}_q)$ denotes its minimum degree.

\textbf{Proof}. Consider an $n$-partite quantum network ${\cal N}_q$ with $\deg({\cal N}_q)=k$, that is, each party shares a bipartite entangled state with at least $k$ parties in ${\cal N}_q$. We need the following lemma, see Appendix H.

\textbf{Lemma 4}. For each pair of parties in ${\cal N}_q$ there are at least $k$ different chain-type subnetworks for connecting them.

From Lemma 4 after ${\cal N}_q$ losing all the particles owned by at most $k-1$ parties, there are at least one chain-type subnetwork connected by any two parties $\textsf{A}_i$ and $\textsf{A}_j$. Hence, after ${\cal N}_q$ passing through the channel ${\cal E}_S(\cdot{})$ with $S=\{\textsf{A}_{j_1}, \cdots, \textsf{A}_{j_{k-1}}\}$, the final subnetwork is 1-connected from Definition 4. By using the recent method \cite{ZDBS}, we can prove that the final state after ${\cal N}_q$ passing through the particle-lose channel ${\cal E}_{S}(\cdot{})$ is genuinely entangled. From Definitions 3 and 5 the robustness-depth of ${\cal N}_q$ is given by $d=\deg({\cal N}_q)-1$. This completes the proof. $\Box$

\textit{Example 6}. Cluster states have inherent network decompositions \cite{Cluster}. As any linear cluster state (chain-type network as shown Fig.\ref{fig-2}(a)) is associated with a specific linear graph with degree no more than 2 \cite{Cluster}, the limited connectedness rules out the robustness against particle loss. The second example is the 2D cluster states \cite{Jozsa2006} as shown Fig.\ref{fig-2}(d), or 3D cluster states which are associated with planar graph or cubic graph, respectively. The robustness-depth of these universal resources is no more than its minimum degree. This implies strong restrictions on the specific computation tasks without error correction. Other examples including the honeycomb networks consisting of GHZ states \cite{Wei2011,GHZ,Luo2018} may be considered similarly.

\section{Losing channel associated with single particles}

So far, each node in entangled networks is regarded as a combined particle. One may consider the lose of partial particles owned by each party, that is, $S$ consists of a single particle in Definition 1. The final state after particle-loss channels may be an $m$-partite state with $m\geq n-|S|$.

\textbf{Definition 6}. Consider an $n$-partite entangled network ${\cal N}_q$ in the state $\rho$ on Hilbert space $\otimes_{i=1}^n\mathbbm{H}_{A_i}$. It is robust entanglement if the output state given by
\begin{eqnarray}
\varrho&=&{\cal E}_{S}(\rho)
\nonumber
\\
&=&\sum_{i\in K_S}E_{i}\rho{}E^\dag_i
\label{eqnss1}
\end{eqnarray}
is $m$-partite entangled for any subset $S$ (consisting of particles) satisfying $|S|\leq N-2$, where $\varrho$ is an $m$-partite state, $K_S$ denotes the Kraus operator decomposition of ${\cal E}_{S}(\cdot{})$, $N$ denotes the number of all the particles and $m\leq N-|S|$.

Different from Definition 1, $S$ in Definition 6 may contain particles owned by more than $n-2$ parties, but cannot contain all the particles owned by $n-1$ parties. Here, the remained particles after ${\cal N}_q$ passing through ${\cal E}_{S}(\cdot{})$ should be owned by different parties. Take triangle network ${\cal N}_q$ shown in Fig.\ref{fig-4}(a) as an example. Suppose that ${\cal N}_q$ consists of three EPR states $|\phi\rangle_{B_1B_2},|\phi\rangle_{B_3B_4}$ and $|\phi\rangle_{B_5B_6}$. One can consider the set $S$ with $|S|=3$ such as $S=\{B_1,B_2,B_3\}$ or $\{B_2,B_3,B_4\}$, or $|S|=4$ such as $S=\{B_1,B_2,B_3,B_4\}$ or $\{B_2,B_3,B_4,B_6\}$. With this definition, we can extend Results 2-4 as follows.

\textbf{Result 5}. The total state of any $k$-independent ($k\geq2$) quantum network ${\cal N}_q$ is not robust in the present model.

For a given $k$-independent quantum network ${\cal N}_q$ with $k\geq2$, there are at least two parties who do not share entanglement after ${\cal N}_q$ passing through the particle-lose channel associated with all the particles owned by other parties. So, the total state of ${\cal N}_q$ is not robust entanglement from Definition 6.

\textbf{Result 6}. Suppose ${\cal N}_q$ is an $n$-partite completely connected network consisting of bipartite entangled pure states and generalized Dicke states. The total state of ${\cal N}_q$ is robust entanglement if one of the following conditions satisfies
\begin{itemize}
\item[(i)] $|S|\leq n-2$;
\item[(ii)] $n-2<|S|\leq N-2$ and the remained subnetwork after ${\cal N}_q$ passing through the channel ${\cal E}_{S}(\cdot{})$ is connected.
\end{itemize}

The proof is easily followed from the connectedness of final networks after passing through particle-lose channels assisted by the recent method \cite{ZDBS}. In fact, from the assumption of Result 6 and Definition 2, each pair of two parties shares at least one bipartite pure entanglement or generalized Dicke state. The connectedness implies that all the final states in Eq.(\ref{eqnss1}) are at least $k$-partite entanglement with $k=n-|S|$ for $|S|\leq n-2$. For $n-2<|S|\leq N-2$, from the assumption we have a connected network after ${\cal N}_q$ passing through the channel ${\cal E}_{S}(\cdot{})$. The proof is completed by using recent method \cite{ZDBS}.

For the case of $n-2<|S|\leq N-2$, Result 6 is different from Result 3. In fact, the remained $m$-partite state $\varrho$ has the entanglement depth less than $m$. Take the triangle network stated above as an example. The remained state is fully separable if $S=\{B_1,B_3,B_5\}$. Hence, the remained subnetwork after passing through the channel ${\cal E}_{S}(\cdot{})$ may be disconnected. In this case, the final state is not entangled. Hence, with the assumption of connectedness, the remained network is entangled. This implies special restrictions on $S$.

\textbf{Definition 7}. An $n$-partite entangled network $\rho$ is robust entanglement with depth $k$ if ${\cal E}_{S}(\rho)$ is entangled for any $S$ with $|S| \leq k-1$, where ${\cal E}_{S}(\cdot)$ is defined in Definition 6.

From Result 4 we get the following result.

\textbf{Result 7}. Suppose that ${\cal N}_q$ is an $n$-partite network consisting of bipartite entangled states. The robustness-depth of ${\cal N}_q$ is given by $d=\deg({\cal N}_q)-1$ for $|S|\leq d-1$, where $\deg({\cal N}_q)$ denotes the minimum degree of ${\cal N}_q$.

The proof is similar to its for Result 4. Different from Result 4, ${\cal N}_q$ may be robust against losing more than $\deg({\cal N}_q)-1$ number of particles. One example is the triangle network stated above, where the final state is entangled for $S=\{B_1,B_2,B_3\}$ while $\deg({\cal N}_q)=2$. However, this does not hold for all the subsets $S$ with $|S|=3$. One example is $S=\{B_1,B_3,B_5\}$. Similarly, there exists one subset $S$ with $|S|=\deg({\cal N}_q)$ such that the final state $\varrho={\cal E}_S(\rho)$ is biseparable for ${\cal N}_q$. For each party $\textsf{A}_j$, one can choose $S$ consisting of all the particles entangled with the party $\textsf{A}_j$.

\section{Conclusions}

It is generally difficult to verify all the outputs of particle-loss channels by using entanglement witnesses \cite{HHH} or Bell inequalities \cite{Sy,BCPS}. The problems can be solved for special states such as generalized symmetric states \cite{DVC,Toth} which have inherent encoding meanings in experiment, and the entangled networks for lang-distant quantum communications or distributed quantum computation. Additionally, the present model is also related to the relaxed absolute maximal entanglement by assuming all the final states after passing particle-losing channels being maximally mixed states.

In conclusion, we have proposed a local model to verify strongly correlated multipartite entanglement robust against particle-loss. This provides an interesting way to characterize single entangled systems or network scenarios going beyond the biseparable model or network local models. It has been used to explore new generic features of multipartite entangled qubit states. This model is useful for characterizing different entangled quantum networks. It is applicable for witnessing the robustness-depth under particle-loss. The present results may highlight further studied on entanglement theory, quantum information processing and measurement-based quantum computation.

\section*{Acknowledgements}

This work was supported by the National Natural Science Foundation of China (Grant Nos.62172341,61772437,12075159), Sichuan Youth Science and Technique Foundation (Grant No.2017JQ0048), Fundamental Research Funds for the Central Universities (Grant No. 2018GF07), Beijing Natural Science Foundation (Grant No.Z190005), Academy for Multidisciplinary Studies, Capital Normal University, Shenzhen Institute for Quantum Science and Engineering, Southern University of Science and Technology (Grant No.SIQSE202001), and the Academician Innovation Platform of Hainan Province.

\appendix

\section{Proof of Lemma 1}

The proof is completed by two subcases.

\textbf{Subcase 1}. $\{|\phi_j^{(i)}\rangle_{A_j}, \forall i\}$ are orthogonal states for all $j$'s.

In this case, the state of $|\Phi\rangle_{A_1\cdots{}A_n}$ in Eq.(\ref{D2}) is equivalent to a generalized GHZ state $\sum_{j=1}^2\sqrt{\lambda_j}|j\cdots{}j\rangle_{A_1\cdots{}A_n}$ under local operations, that is,
\begin{eqnarray}
(\otimes_{j=1}^nW_{A_j})|\Phi\rangle_{A_1\cdots{}A_n}=
\sqrt{p_1}|0\cdots0\rangle+\sqrt{1-p_1}|1\cdots1\rangle
\nonumber
\\
\label{D5}
\end{eqnarray}
where $W_{A_j}$ are unitary operations on the qubit $A_j$ defined by
\begin{eqnarray}
W_{A_j}:|\phi_k^{(j)}\rangle_{A_j}\mapsto |k\rangle_{A_j}
\end{eqnarray}
for $j=1, \cdots, n$.

\textbf{Subcase 2}. $\{|\phi_j^{(k)}\rangle_{A_j}, \forall k\}$ are not orthogonal states for some $k$'s.

Define the local unitary operation $W_{A_k}$ as
\begin{eqnarray}
W_{A_k}:|\phi_j^{(k)}\rangle\mapsto |j\rangle, j=0, 1
\label{D7}
\end{eqnarray}
for $k=1, \cdots, n$. Since local unitary operations do not change the entanglement, from Eq.(\ref{D2}) it is sufficient to consider the following state
\begin{eqnarray}
|\Psi\rangle_{A_1\cdots{}A_n}&:=&(\otimes_{i\not=2} W_{A_i})|\Phi\rangle
\nonumber
\\
&=&\sqrt{1-p_0}|1\rangle_{A_1}
|1\rangle_{A_2}(\otimes_{j=3}^n|\psi_j\rangle_{A_j})
\nonumber
\\
&&
+\sqrt{p_0}|0\cdots0\rangle_{A_1\cdots{}A_n}
\label{D8}
\end{eqnarray}

Without changing of notations, assume that $\{|\phi_2^{(i)}\rangle_{A_2}, \forall i\}$ are orthogonal states from the orthogonality of $|\Phi_0\rangle$ and $|\Phi_1\rangle$ in Eq.(\ref{D2}), and $|\psi_j\rangle:=W_{A_j}|\phi_j^{(2)}\rangle_{A_j}$ for $j=3, \cdots, n$. For the simplicity, suppose that
\begin{eqnarray}
|\psi_j\rangle=a_{0j}|0\rangle+a_{1j}|1\rangle
\label{D9}
\end{eqnarray}
for $j=3, \cdots, n$.

Let $|\Psi\rangle_{A_1\cdots{}A_n}$ in Eq.(\ref{D8}) pass through the particle-lose channels ${\cal{} E}_{A_3\cdots{}A_n}(\cdot)$. The remained state is given by
\begin{eqnarray}
    \rho_{(A_3\cdots{}A_n)}&=&{\cal{} E}_{A_3\cdots{}A_n}(|\Psi\rangle_{A_1\cdots{}A_n}\langle \Psi|)
    \nonumber
    \\
    &=&c |\hat{\phi}\rangle_{A_1A_2}\langle \hat{\phi}|+(1-c)|11\rangle_{A_1A_2}\langle 11|,
    \label{D10}
\end{eqnarray}
where $|\hat{\phi}\rangle_{A_1A_2}$ is given by
\begin{eqnarray}
|\hat{\phi}\rangle_{A_1A_2}=\frac{1}{\sqrt{c}}(\sqrt{p_0}|00\rangle+\sqrt{1-p_0}\prod_{j=3}^na_{0j}|11\rangle)
\end{eqnarray}
and $c=p_0+(1-p_0)\prod_{j=3}^na_{0j}$. $\rho_{(A_{3}\cdots{}A_n)}$ in Eq.(\ref{D10}) is a bipartite entanglement from the PPT criterion \cite{PPT} for any $\prod_{j=3}^na_{0j}\not=0$. This contradicts to the assumption that $\rho_{(A_i)}$ is fully separable for any $i$. Hence, there is $j$ such that $a_{0j}=0$. Take $j=3$ for an example. From Eq.(\ref{D8}) it follows that
\begin{eqnarray}
|\Psi\rangle&=&\sqrt{1-p_0}|111\rangle_{A_1A_2A_3}\otimes_{j=4}^n|\psi_j\rangle_{A_j}
\nonumber
\\
&&+\sqrt{p_0}|0\cdots{}0\rangle_{A_1\cdots{}A_n}
\label{D11}
\end{eqnarray}

Let $|\Psi\rangle$ in Eq.(\ref{D11}) pass through the particle-lose channel ${\cal{} E}_{A_4\cdots{}A_n}(\cdot)$. The remained state is given by
\begin{eqnarray}
    \rho_{(A_4\cdots{}A_n)}&=&{\rm Tr}_{A_4\cdots{}A_n}(|\Psi\rangle\langle \Psi|)
    \nonumber
    \\
    &=&c'|\tilde{\phi}\rangle_{A_1A_2A_3}\langle \tilde{\phi}|+(1-c')|111\rangle_{A_1A_2A_3}\langle 111|,
\nonumber
\\
    \label{D12}
\end{eqnarray}
where $|\tilde{\phi}\rangle_{A_1A_2A_3}$ is given by
\begin{eqnarray}
|\tilde{\phi}\rangle_{A_1A_2A_3}=\frac{1}{\sqrt{c}}(\sqrt{p_0}|000\rangle
+\sqrt{1-p_0}\prod_{j=4}^na_{0j}|111\rangle)
\nonumber
\\
\end{eqnarray}
and $c'=p_0+(1-p_0)\prod_{j=4}^na_{0j}$.

Note that $\rho_{(A_4\cdots{}A_n)}$ is a genuinely tripartite entanglement \cite{Sy} for any $\prod_{j=4}^na_{0j}\not=0$, where $|\tilde{\phi}\rangle_{A_1A_2A_3}$ is genuinely tripartite entangled \cite{GHZ}. This contradicts to the assumption that $\rho_{(A_i)}$ is fully separable for any $i$. Hence, there is an integer $j$ with $a_{0j}=0$. Take $j=4$ for an example. From Eq.(\ref{D11}) we get
\begin{eqnarray}
|\Psi\rangle_{A_1\cdots{}A_n}&=&\sqrt{1-p_0}\otimes_{s=1}^4|1\rangle_{A_s}
\otimes_{j=5}^n|\psi_j\rangle_{A_j}
\nonumber\\
&&+\sqrt{p_0}|0\cdots0\rangle.
\label{D13}
\end{eqnarray}

The procedure stated above can be iteratively performed for $j=5, \cdots, n$. This implies that
\begin{eqnarray}
|\Psi\rangle_{A_1\cdots{}A_n}=\sqrt{p_0}|0\cdots0\rangle+\sqrt{1-p_0}|1\cdots1\rangle
\label{D14}
\end{eqnarray}
under local operations. Hence, from Eq.(\ref{D8}), the state of $|\Phi\rangle_{A_1\cdots{}A_n}$ is equivalent to a generalized $n$-partite GHZ state under local unitary operations. This completes the proof for the subcase 2.

\section{Proof of Lemma 2}

By using the normal form \cite{Kraus2009} of $|\Phi\rangle_{A_1\cdots{}A_n}$, from Eq.(\ref{D2}) assume that
\begin{eqnarray}
|\Phi\rangle=\sqrt{p_0}|0\cdots0\rangle_{A_1\cdots{}A_n}+\sqrt{1-p_0}
|1\rangle_{A_1} |\hat{\Phi}\rangle_{A_2\cdots{}A_n}
\label{D15}
\end{eqnarray}
where $|\hat{\Phi}\rangle$ is an $n-1$-partite state of qubits $A_2,\cdots, A_n$, which is orthogonal to the state $|0\cdots0\rangle_{A_2\cdots{}A_n}$.

In what follows, the main goal is to prove $|\hat{\Phi}\rangle=|1\cdots1\rangle_{A_2\cdots{}A_n}$. Consider the subsystem after $|\Phi\rangle$ passing through the particle-lose channel ${\cal E}_{A_1}(\cdot)$ as follows
\begin{eqnarray}
\rho_{(A_1)}=p_0\otimes_{j=2}^n|0\rangle\langle 0|_{A_j}+(1-p_0) |\hat{\Phi}\rangle_{A_2\cdots{}A_n}\langle \hat{\Phi}|
\label{D16}
\end{eqnarray}
Note that $\rho_{(A_1)}$ is fully separable. Assume that \begin{eqnarray}
|\hat{\Phi}\rangle_{A_2\cdots{}A_n}=\otimes_{j=2}^n|\psi_j\rangle_{A_j}.
\end{eqnarray}
Moreover, $|\hat{\Phi}\rangle_{A_2\cdots{}A_n}$ and $|0\cdots0\rangle_{A_2\cdots{}A_n}$ are orthogonal states from Eq.(\ref{D15}). From Eqs.(\ref{D8})-(\ref{D14}), we get that $|\hat{\Phi}\rangle=|1\cdots1\rangle_{A_2\cdots{}A_n}$. So, the state of $|\Phi\rangle$ in Eq.(B1) is equivalent to a generalized GHZ state in Eq.(\ref{eqnaa1}) under local unitary operations. This completes the proof.

\section{Proof of Lemma 3}

We prove that $p=\frac{1}{2}$ and $|\Phi_0\rangle_{A_2\cdots A_n}$ and $|\Phi_1\rangle_{A_2\cdots A_n}$ have special decompositions, that is, the qubits $A_1$ and $A_2$ are symmetric. From the Schmidt decomposition, by using the normal form \cite{Kraus2009} suppose that
\begin{eqnarray}
|\Phi_0\rangle=\alpha_0|0\rangle_{A_2}|\phi_0\rangle_{A_3\cdots A_n}+\beta_0
|1\rangle_{A_2}|\phi_1\rangle_{A_3\cdots A_n}
\label{D17}
\end{eqnarray}
where $|\phi_0\rangle$ and $|\phi_1\rangle$ are orthogonal states.

Define $W$ as an $n-2$-partite unitary operation given by
\begin{eqnarray}
 W:|\phi_i\rangle_{A_3\cdots{}A_n}\mapsto |i\cdots i\rangle_{A_3\cdots{}A_n}, i=0,1.
 \label{D18}
\end{eqnarray}
This can be extended to general unitary operation on Hilbert space $\otimes_{j=3}^n\mathbb{H}_{A_j}$. From Eqs.(\ref{D17}) and (\ref{D18}), we get that
\begin{eqnarray}
(\mathbbm{1}_{A_2}\otimes{}W)|\Phi_0\rangle_{A_2\cdots{}A_n}&=&
\alpha_0|0\cdots0\rangle+\beta_0
|1\cdots1\rangle,
\nonumber
\\
(\mathbbm{1}_{A_2}\otimes{}W)|\Phi_1\rangle_{A_2\cdots{}A_n}&=&\alpha_1|0\cdots0\rangle+\beta_1
|1\cdots1\rangle
\nonumber
\\
&&+\delta|\hat{\Phi}_1\rangle_{A_2\cdots{}A_n},
\label{D19}
\end{eqnarray}
where $|\hat{\Phi}_1\rangle_{A_2\cdots{}A_n}$ and $\mathbbm{1}_{A_2}\otimes{}W|\Phi_0\rangle_{A_2\cdots{}A_n}$ are orthogonal states. Assume that $|\hat{\Phi}_1\rangle_{A_2\cdots{}A_n}$ is on the subspace spanned by all the $n-1$ qubit states $|i_2\cdots i_{n}\rangle_{A_2\cdots{}A_n}$ except for $|0\cdots0\rangle_{A_2\cdots{}A_n}$ and $|1\cdots 1\rangle_{A_2\cdots{}A_n}$.

In what follows, we prove $\delta=0$ and $p=1/2$. In fact, consider the bipartition of $\{A_2\}$ and $\{A_3, \cdots, A_n\}$. $(\mathbbm{1}_{A_2}\otimes{}W_{A_3\cdots A_n})\rho_{(A_1)}(\mathbbm{1}_{A_2}\otimes W^\dag_{A_3\cdots A_n})$ is biseparable because $\rho_{(A_1)}$ is fully separable, where $\mathbbm{1}_{A_2}$ is the identity operator on qubit $A_2$. Hence, by using the PPT criterion \cite{PPT} we get
\begin{eqnarray}
\hat{\rho}^T_{A_2}\geq 0,
\label{D20}
\end{eqnarray}
where $\hat{\rho}^T_{A_2}$ is defined by $\hat{\rho}^T_{A_2}=(x_{ij,k\ell})$ with $x_{ij,k\ell}=y_{kj,i\ell}$ and $\hat{\rho}_{(A_1)}=(y_{ij,k\ell})$ in terms of the bipartition $\{A_2\}$ and $\{A_3, \cdots, A_n\}$.

Consider the principal minor $D_{0\overline{1}}$ of $\hat{\rho}^T_{A_2}$ defined by the matrix basis $\{|00\rangle\langle \overline{1}\,\overline{1}|, |0\overline{1}\rangle\langle \overline{1}0|, |\overline{1}0\rangle\langle 0\overline{1}|, |\overline{1}\,\overline{1}\rangle\langle 00|\}$ with $\overline{1}=1\cdots 1$ ($n-2$ number of 1). It follows that
\begin{eqnarray}
p\alpha_0\beta_0+(1-p)\alpha_1\beta_1=0.
\label{C21}
\end{eqnarray}
Combined with $\alpha_0\alpha_1+\beta_0\beta_1=0$, that is, the orthogonality of $|\Phi_i\rangle$'s, we get that
\begin{eqnarray}
\alpha_1=\pm\alpha_0\gamma,\beta_1=\mp\beta_0\gamma, p=\frac{1}{1+\gamma^2}.
\label{D22}
\end{eqnarray}

Now, we will prove that $\gamma=1$ or $\delta=0$, that is, there does not exist $|\hat{\Phi}_1\rangle$ in Eq.(\ref{D19}). The proof is completed by contradiction. In fact, consider the principal minor of $\hat{\rho}^T_{A_2}$ defined on the subspace associated with $|\hat{\Phi}_1\rangle$. Since $|\hat{\Phi}_1\rangle$ and $\alpha_1|0\cdots0\rangle_{A_2\cdots{}A_n}+\beta_1
|1\cdots1\rangle_{A_2\cdots{}A_n}$ are states on different subspaces, the minor depends only on the state of $|\hat{\Phi}_1\rangle_{A_2\cdots{}A_n}$. From Eq.(\ref{D20}), it follows that $\gamma=0$, or $|\hat{\Phi}_1\rangle_{A_2\cdots{}A_n}$ is biseparable \cite{Sy} in terms of the bipartition $\{A_2\}$ and $\{A_3, \cdots, A_n\}$, that is,  $|\hat{\Phi}_1\rangle=|\phi\rangle_{A_2}|\hat{\Phi}_2\rangle_{A_3\cdots A_n}$. In what follows, it only needs to prove the second case.

Generally, suppose that
\begin{eqnarray}
|\hat{\Phi}_1\rangle_{A_2\cdots{}A_n}=|\phi\rangle_{A_2}|\hat{\Phi}_2\rangle_{A_3\cdots A_n}.
\label{D23}
\end{eqnarray}
Note that $|\hat{\Phi}_1\rangle_{A_2\cdots{}A_n}$ does not contain the terms of $|0\rangle^{\otimes n-1}$ and $|1\rangle^{\otimes n-1}$. There are three subcases.

\begin{itemize}
\item[(i)] $|\phi\rangle_{A_2}=|0\rangle$. We get $|\hat{\Phi}_2\rangle_{A_3\cdots A_n}$ does not contain the terms of $|0\rangle^{\otimes n-2}$. Suppose that
\begin{eqnarray}
|\hat{\Phi}_2\rangle_{A_3\cdots A_n}=\alpha_2|\hat{\Phi}_3\rangle_{A_3\cdots A_n}+\beta_2|1\cdots1\rangle_{A_3\cdots A_n}.
\label{D24}
\end{eqnarray}
 Define $H$ as an $n-2$-partite unitary operation $H$ given by
\begin{eqnarray}
 H: &&|i\cdots{}i\rangle_{A_3\cdots A_n}\mapsto |i\cdots{}i\rangle_{A_3\cdots A_n},
 \nonumber\\
 &&|\hat{\Phi}_3\rangle_{A_3\cdots A_n}\mapsto |0\rangle_{A_3}|1\cdots 1\rangle_{A_4\cdots A_n}.
 \label{D25}
\end{eqnarray}
From Eqs.(\ref{D19}), (\ref{D22}) and (\ref{D25}), it follows that
\begin{eqnarray}
   (\mathbbm{1}_{A_2}+HW)|\Phi_0\rangle&=&(\mathbbm{1}_{A_2}\otimes W)|\Phi_0\rangle,
   \nonumber \\
   (\mathbbm{1}_{A_2}+HW)|\Phi_1\rangle&=&\alpha_1|0\cdots 0\rangle_{A_2\cdots A_n}
    \nonumber\\
 &&+\beta_1|1\cdots 1\rangle_{A_2\cdots A_n}
    \nonumber\\
 &&
+\delta \alpha_2|0\rangle_{A_2}|1\cdots 1\rangle_{A_3\cdots A_n}
   \nonumber\\
 &&+\delta \beta_2|00\rangle_{A_2A_3}|1\cdots1\rangle_{A_3\cdots A_n}.
 \nonumber\\
\label{D26}
\end{eqnarray}
Now, consider the remained state $\rho_{(A_{1}A_4
\cdots{}A_n)}$  as
\begin{eqnarray}
&&\rho_{(A_{1}A_4
\cdots{}A_n)}
    \nonumber\\
&=&{\cal{}E}_{A_4\cdots A_n}((\mathbbm{1}_{A_2}+HW)\rho_{(A_1)}(\mathbbm{1}_{A_2}+HW)^\dag)
\nonumber
\\
&=& (p\alpha_0^2+(1-p)\alpha_1^2)|00\rangle_{A_2A_3}\langle 00|
    \nonumber\\
 && +p\beta_0^2|11\rangle_{A_2A_3}\langle 11|
+(1-p)\tau|\psi\rangle_{A_2A_3}\langle \psi|,
\label{D27}
\end{eqnarray}
where $|\psi\rangle_{A_2A_3}$ is given by
\begin{eqnarray}
|\psi\rangle_{A_2A_3}=\frac{1}{\tau}(\delta\beta_2|00\rangle+\delta\alpha_2|01\rangle+\beta_1|11\rangle)
\end{eqnarray}
and $\tau=\sqrt{\delta^2+\beta_1^2}$. Note that $\rho_{(A_{1}A_4
\cdots{}A_n)}$ should be separable because $\rho_{(A_1)}$ is fully separable, where $H$ and $W$ are not performed on qubit $A_2$. Hence, it follows that $\delta=0$ or $\beta_2=0$ by using the PPT criterion \cite{PPT}, that is, $|\psi\rangle_{A_2A_3}$ should be separable. From Eq.(\ref{D24}) it yields to
\begin{eqnarray}
|\hat{\Phi}_2\rangle_{A_3\cdots A_n}=\alpha_2|\hat{\Phi}_3\rangle.
\label{D28}
\end{eqnarray}

Similarly, suppose that $H$ in Eq.(\ref{D25}) is replaced by the following operation
\begin{eqnarray}
H: &&|\hat{\Phi}_3\rangle_{A_3\cdots A_n}\mapsto |1\rangle_{A_3}|0\cdots 0\rangle_{A_4\cdots A_n},
\nonumber\\
&&|i\cdots i\rangle_{A_3\cdots A_n}\mapsto |i\cdots i\rangle_{A_3\cdots A_n}.
\label{D29}
\end{eqnarray}
We can prove that $\alpha_2=0$ or $\delta=0$ by using the separability of the reduced density matrix $\rho_{(A_{1}A_4\cdots{}A_n)}={\cal E}_{A_4\cdots A_n}((\mathbbm{1}_{A_2}+HW)\rho_{(A_1)}(\mathbbm{1}_{A_2}+HW)^\dag)$. So, we get $\delta=0$.
\item[(ii)]  $|\phi\rangle_{A_2}=|1\rangle$. We get that $|\hat{\Phi}_2\rangle_{A_3\cdots A_n}$ does not contain the terms of $|1\cdots1\rangle_{A_3\cdots A_n}$. Suppose that
    \begin{eqnarray}
    |\hat{\Phi}_2\rangle=\alpha_2|\hat{\Phi}_3\rangle_{A_3\cdots{}A_n}+\beta_2|0\cdots0\rangle_{A_3\cdots A_n}.
    \label{D30}
    \end{eqnarray}
    Similar to Eqs.(\ref{D25})-(\ref{D29}), we get $\delta=0$.
\item[(iii)] $|\phi\rangle_{A_2}=\alpha_2|0\rangle+\beta_2|1\rangle$. In this case, $|\hat{\Phi}_2\rangle_{A_3\cdots A_n}$ does not contain the terms of $|0\cdots0\rangle_{A_3\cdots A_n}$ and $|1\cdots1\rangle_{A_3\cdots A_n}$. Define $H$ as a unitary operation given by
    \begin{eqnarray}
    H:
&&|\hat{\Phi}_2\rangle_{A_3\cdots A_n}\mapsto |1\rangle_{A_3}|0\cdots0\rangle_{A_4\cdots A_n},
    \nonumber
    \\
 &   &|i\cdots i\rangle_{A_3\cdots A_n}\mapsto |i\cdots i\rangle_{A_3\cdots A_n}.
   \label{D31}
     \end{eqnarray}
   From Eqs.(\ref{D19}), (\ref{D22}) and (\ref{D31}) we get that
    \begin{eqnarray}
   (\mathbbm{1}_{A_2}+HW)|\Phi_0\rangle_{A_2\cdots{}A_n}&=&(\mathbbm{1}_{A_2}+W)|\Phi_0\rangle,
\end{eqnarray}
and
\begin{eqnarray}
 &&(\mathbbm{1}_{A_2}+HW)|\Phi_1\rangle_{A_2\cdots{}A_n}
 \nonumber \\
 &=&\alpha_1|0\cdots0\rangle_{A_2\cdots{}A_n}
 +\beta_1|1\cdots 1\rangle_{A_2\cdots{}A_n}
    \nonumber\\
 && +\delta (\alpha_2|0\rangle+\beta_2|1\rangle)_{A_2}|1\rangle_{A_3}
 |0\cdots0\rangle_{A_4\cdots{}A_n}.
\label{D32}
    \end{eqnarray}
Consider the remained state $\rho_{(A_1\cdots{}A_4\cdots{}A_n)}$ after passing through the particle-lose channel ${\cal E}_{A_4\cdots A_n}$ as
\begin{eqnarray}
&&\rho_{(A_1\cdots{}A_4\cdots{}A_n)}
\nonumber\\
&=&{\cal E}_{A_4\cdots A_n}(\mathbbm{1}_{A_2}+HW)\rho_{(A_1)}(\mathbbm{1}_{A_2}+HW)^\dag)
\nonumber
\\
&=&(p\beta_0^2+(1-p)\beta_1^2)|11\rangle_{A_2A_3}\langle 11|
    \nonumber\\
 && p\alpha_0^2|00\rangle_{A_2A_3}\langle 00| +(1-p)\tau|\psi'\rangle_{A_2A_3}\langle \psi'|.
\label{D33}
\end{eqnarray}
where $|\psi'\rangle_{A_2A_3}$ is given by
\begin{eqnarray}
|\psi'\rangle=\frac{1}{\sqrt{\tau'}}(\alpha_1|00\rangle
+\delta (\alpha_2|0\rangle+\beta_2|1\rangle)_{A_2}|1\rangle_{A_3})
\end{eqnarray}
 and $\tau'=\sqrt{\alpha_1^2+\delta^2}$. $\rho_{(A_{1}A_4\cdots{}A_n)}$ should be separable since $\rho_{(A_1)}$ is fully separable, where $H$ and $W$ are not performed on qubit $A_2$. Hence, it follows that $\delta=0$ or $\alpha_1=0$ by using the PPT criterion \cite{PPT}. However, from the assumption in Case (iii), we have $\alpha_1\not=0$. It follows that $\delta=0$.
\end{itemize}

Hence, we have proved $\delta=0$. Combining with Eqs.(\ref{D17}) and (\ref{D19}), $|\Phi_i\rangle$ in Eq.(\ref{D2}) is written into
\begin{eqnarray}
|\Phi_1\rangle=\beta_0|0\rangle_{A_2}|\phi_0\rangle_{A_3\cdots A_n}-\alpha_0
|1\rangle_{A_2}|\phi_1\rangle_{A_3\cdots A_n}
\label{D34}
\end{eqnarray}
From Eqs.(\ref{D2}), (\ref{D17}) and (\ref{D34}), it follows that
\begin{eqnarray}
|\Phi\rangle&=&\frac{1}{2}(\alpha_0|0\rangle\pm\beta_0|1\rangle)_{A_1}|0\rangle_{A_2}|\phi_0\rangle_{A_3\cdots A_n}
    \nonumber\\
 && +\frac{1}{2}(\beta_0|0\rangle\mp\alpha_0|1\rangle)_{A_1}|1\rangle_{A_2}|\phi_1\rangle_{A_3\cdots A_n}
\label{D35}
\end{eqnarray}
Define $U_{A_1}^\pm$ as local unitary operations given by
\begin{eqnarray}
U_{A_1}^\pm:&&\alpha_0|0\rangle_{A_1}\pm\beta_0|1\rangle_{A_1}\mapsto |0\rangle,
\nonumber
\\
&&\beta_0|0\rangle_{A_1}\mp\alpha_0|1\rangle_{A_1}\mapsto |1\rangle.
\end{eqnarray}
It follows that
\begin{eqnarray}
(U_{A_1}^\pm+\mathbbm{1}_{A_2\cdots{}A_n})
|\Phi\rangle&=&\frac{1}{2}(|00\rangle_{A_1A_2}|\phi_0\rangle_{A_3\cdots{}A_n}
\nonumber\\
&&+|11\rangle_{A_1A_2}|\phi_1\rangle_{A_3\cdots{}A_n})
\label{D35}
\end{eqnarray}
where $\mathbbm{1}_{A_2\cdots{}A_n}$ denotes the identity operator on qubits $A_2, \cdots, A_n$. So, the qubits $A_1$ and $A_2$ in $|\Phi\rangle$ are symmetric under local unitary operations.

\section{The star-convexity of ${\cal B}_d$}

Consider Hilbert space $\mathbb{H}:=\mathbb{H}_{A_1}\otimes \cdots \otimes\mathbb{H}_{A_n}$. The goal is to construct a new state $\rho=p\rho_1+(1-p)\rho_2$ such that $\rho\not\in {\cal B}_d$ for some $p$, and $\rho_1,\rho_2\in {\cal B}_d$ on $\mathbb{H}$. Define
\begin{eqnarray}
\rho_1&=&|\Phi_1\rangle_{A_1A_2A_3} \langle\Phi_1|,
\nonumber\\
\rho_2&=&|\Phi_1\rangle_{A_1A_2A_3} \langle\Phi_2|
\label{A1}
\end{eqnarray}
with
\begin{eqnarray}
|\Phi_1\rangle=\frac{1}{2}(|000\rangle+|011\rangle+|120\rangle+|131\rangle),
\nonumber\\
|\Phi_2\rangle=\frac{1}{2}(|000\rangle+|101\rangle+|210\rangle+|311\rangle).
\label{A2}
\end{eqnarray}
It is easy to prove that $|\Phi_1\rangle$ and $|\Phi_2\rangle$ are genuinely tripartite entanglement \cite{Sy}. Moreover, let $\rho_{(A_i)}^{(j)}$ be the output state of $\rho_j$ after passing through the particle-lose channel ${\cal{} E}_{A_i}(\cdot)$, $i=1, \cdots, 3; j=1, 2$, where the density matrices of $\rho_{(A_i)}^{(j)}$ are defined by
\begin{eqnarray}
\rho_{(A_1)}^{(1)}&=&\rho_{(A_2)}^{(2)}=\frac{1}{2}|\phi_{00}\rangle\langle \phi_{00}|+\frac{1}{2}|\psi_{20}\rangle\langle \psi_{20}|,
\nonumber\\
\rho_{(A_2)}^{(1)}&=&\rho_{(A_1)}^{(2)}=\frac{1}{4}(|00\rangle\langle 00|+|01\rangle\langle 01|
    \nonumber\\
 && +|10\rangle\langle 10|+|11\rangle\langle 11|),
\nonumber\\
\rho_{(A_3)}^{(1)}&=&\rho_{(A_3)}^{(2)}=\frac{1}{2}|\psi_{12}\rangle\langle \psi_{12}|+ \frac{1}{2}|\psi_{01}\rangle\langle \psi_{01}|,
\label{A3}
\end{eqnarray}
and
\begin{eqnarray*}
|\phi_{00}\rangle&=&\frac{1}{\sqrt{2}}(|00\rangle+|11\rangle),
|\psi_{20}\rangle=\frac{1}{\sqrt{2}}(|20\rangle+|31\rangle),
\\
|\psi_{12}\rangle&=&\frac{1}{\sqrt{2}}(|00\rangle+|12\rangle),
|\psi_{01}\rangle=\frac{1}{\sqrt{2}}(|01\rangle+|13\rangle).
\end{eqnarray*}
Note that $|\psi_{ij}\rangle$'s are equivalent to the EPR state \cite{EPR} under local unitary operations. Moreover, $\rho_{(A_1)}^{(1)},\rho_{(A_2)}^{(2)}, \rho_{(A_3)}^{(1)}$ and $\rho_{(A_3)}^{(2)}$ are bipartite entangled states, where $|\psi_{ij}\rangle$'s are defined on different subspaces which rule out separable decompositions. Instead, $\rho_{(A_2)}^{(1)}$ and $\rho_{(A_1)}^{(2)}$ are separable. So, we get $\rho_1, \rho_2\in {\cal B}_d$.

Now, define
\begin{eqnarray}
\varrho=\frac{1}{2}\rho_1+\frac{1}{2}\rho_2.
\label{A4}
\end{eqnarray}
It follows from Eq.(\ref{A3}) that
\begin{eqnarray}
\varrho_{(A_1)}&=&\varrho_{(A_2)}
\nonumber
\\
&=&\frac{1}{4}|\phi_{00}\rangle\langle \phi_{00}|+\frac{1}{4}|\psi_{20}\rangle\langle \psi_{20}|+\frac{1}{8}(|00\rangle\langle 00|
    \nonumber\\
 && +|01\rangle\langle 01|+|10\rangle\langle 10|+|11\rangle\langle 11|)
\nonumber\\
\varrho_{(A_3)}&=&\frac{1}{2}|\psi_{12}\rangle\langle \psi_{12}|+ \frac{1}{2}|\psi_{01}\rangle\langle \psi_{01}|.
\label{A5}
\end{eqnarray}
Since $|\psi_{20}\rangle$ is defined on specific subspace spanned by $\{|20\rangle, |31\rangle\}$, which is different from the associated subspace of $\frac{1}{4}|\phi_{00}\rangle\langle \phi_{00}|+\frac{1}{8}(|00\rangle\langle 00|+|01\rangle\langle 01|+|10\rangle\langle 10|+|11\rangle\langle 11|)$. $\varrho_{(A_1)}$ and $\varrho_{(A_2)}$ are bipartite entangled states. So, $\varrho_{(A_i)}$'s are bipartite entangled. Moreover, $\varrho$ is a genuinely tripartite entanglement \cite{PPT}, where $\rho_1$ can be decomposed into the chain-shape network as shown in Fig.2(a) \cite{Luo2018}. It means that $\rho\not \in {\cal B}_d$. Hence, ${\cal B}_d$ is not convex.

The star convexity of ${\cal B}_d$ can be followed by choosing the maximally mixed state $\rho_{0}=\frac{1}{2^n}\mathbbm{1}$ on Hilbert space $\otimes_{i}\mathbb{H}_{A_i}$ as the center point. For any state $\rho\in {\cal B}_d$, it is easy to prove that $p\rho+(1-p)\rho_0\in {\cal B}_d$ for any $p\in (0,1)$. This completes the proof.

\section{The useless of CHSH inequality}

We prove that the CHSH inequality \cite{CHSH} cannot be applied for verifying the strong nonlocality of the generalized W state \cite{DVC,Fei,Cheng} as
\begin{eqnarray}
|W\rangle_{A_1A_2A_3}=\alpha|001\rangle+\beta|010\rangle+\gamma|100\rangle
\label{K2}
\end{eqnarray}
with $\alpha^2+\beta^2+\gamma^2=1$. The proof is completed for three states $\rho_{(A_3)}$, $\rho_{(A_2)}$ and $\rho_{(A_1)}$. In fact, from Eq.(\ref{K2}) it follows that
\begin{eqnarray}
&&\rho_{(A_3)}=\alpha^2|00\rangle_{A_1A_2}\langle 00|+(1-\alpha^2)|\psi_1\rangle_{A_1A_2}\langle \psi_1|,
\nonumber\\
&&\rho_{(A_2)}=\beta^2|00\rangle_{A_1A_3}\langle 00|+(1-\beta^2)|\psi_2\rangle_{A_1A_3}\langle \psi_2|,
\nonumber\\
&&\rho_{(A_1)}=\gamma^2|00\rangle_{A_2A_3}\langle 00|+(1-\gamma^2)|\psi_3\rangle_{A_2A_3}\langle \psi_3|,
\label{K3}
\end{eqnarray}
where $|\psi_i\rangle$ are given by  $|\psi_1\rangle=\frac{1}{\sqrt{1-\alpha^2}}(\beta|01\rangle+\gamma|10\rangle)$, $|\psi_2\rangle=\frac{1}{\sqrt{1-\beta^2}}(\alpha|01\rangle+\gamma|10\rangle)$ and
$|\psi_3\rangle=\frac{1}{\sqrt{1-\gamma^2}}(\alpha|01\rangle+\beta|10\rangle)$.

\begin{figure}
\begin{center}
\resizebox{240pt}{180pt}{\includegraphics{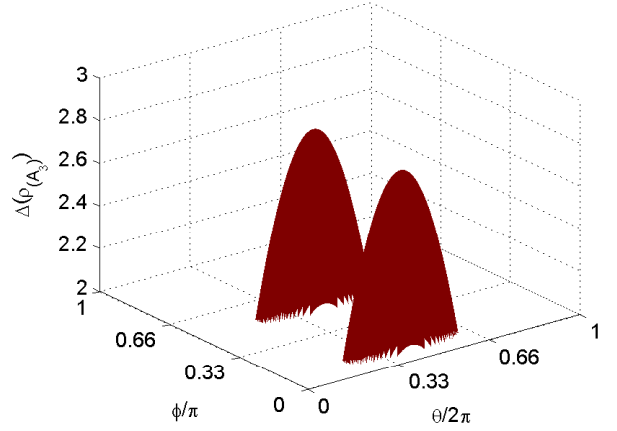}}
\end{center}
\caption{\small (Color online) The violation of CHSH inequality for the state of $\rho_{(A_3)}$. Here, $\Delta(\rho_{(A_3)})=4\sqrt{2}\beta\gamma-2$,  $\alpha=\sin\theta\cos\phi,\beta=\sin\theta\sin\phi$,  and $\gamma=\cos\theta$. The maximal violation of $2.8284$ is unavailable because of $\alpha=0$.}
\label{fig-S1}
\end{figure}

There are two cases to achieve the maximal violation of the CHSH inequality \cite{CHSH}. One is from local measurements on the $x$-$y$ plane of Pauli sphere. The other is from local measurements on $x$-$z$ plane of Pauli sphere. Specially, take local observables $A_1=X$ and $A_2=Y$ for the first party, $B_1=\cos\theta{}X+i\sin\theta{}Y$ and $B_2=\cos\theta{}X-i\sin\theta{}Y$ for the second party. For the state of $\rho_{(A_3)}$, it follows that
\begin{eqnarray}
 && \langle A_1B_1\rangle+\langle A_1B_2\rangle+\langle A_2B_1\rangle-\langle A_2B_2\rangle
 \nonumber\\
&=&4\cos\theta \beta\gamma +4\sin\theta \beta\gamma
\nonumber
\\
&=&
4\sqrt{2}\beta\gamma
\label{K4}
\end{eqnarray}
when $\theta=\frac{\pi}{4}$. This means that the statistics generated from local measurements on $\rho_{(A_3)}$ violates the CHSH inequality \cite{CHSH} if $|\beta\gamma|>\frac{1}{2\sqrt{2}}$, as shown in Fig.\ref{fig-S1}.

\begin{figure}
\begin{center}
\resizebox{240pt}{200pt}{\includegraphics{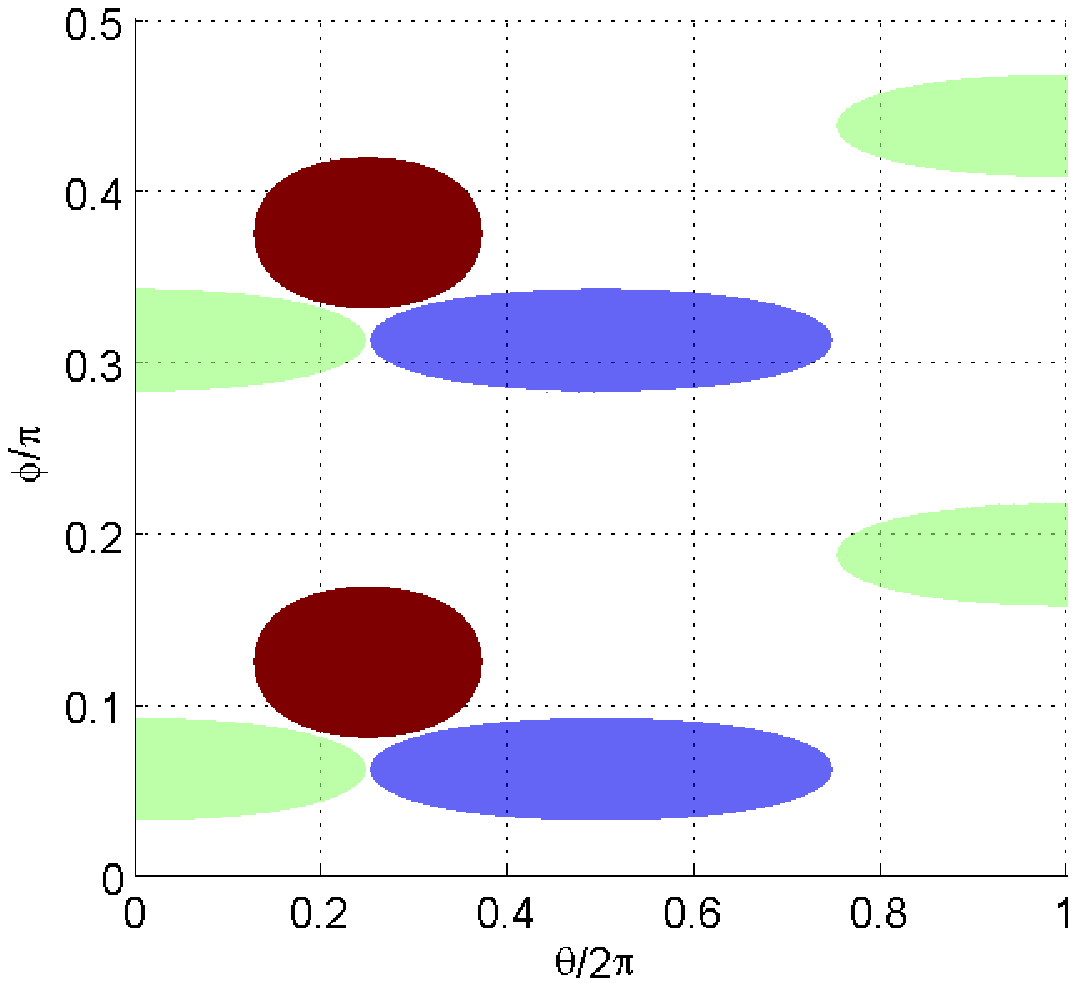}}
\end{center}
\caption{\small (Color online) The violations of the CHSH inequality with observablei from the $x$-$y$ plane of Pauli sphere. Here, the red region is for $\rho_{(A_3)}$. The blue region is for $\rho_{(A_2)}$. The green region is for $\rho_{(A_1)}$. It shows that there is no choice of $\alpha, \beta$ and $\gamma$ such that its statistics of $\rho_{(A_j)}$'s violate the CHSH inequality.}
\label{fig-S3}
\end{figure}

\begin{figure}
\begin{center}
\resizebox{240pt}{200pt}{\includegraphics{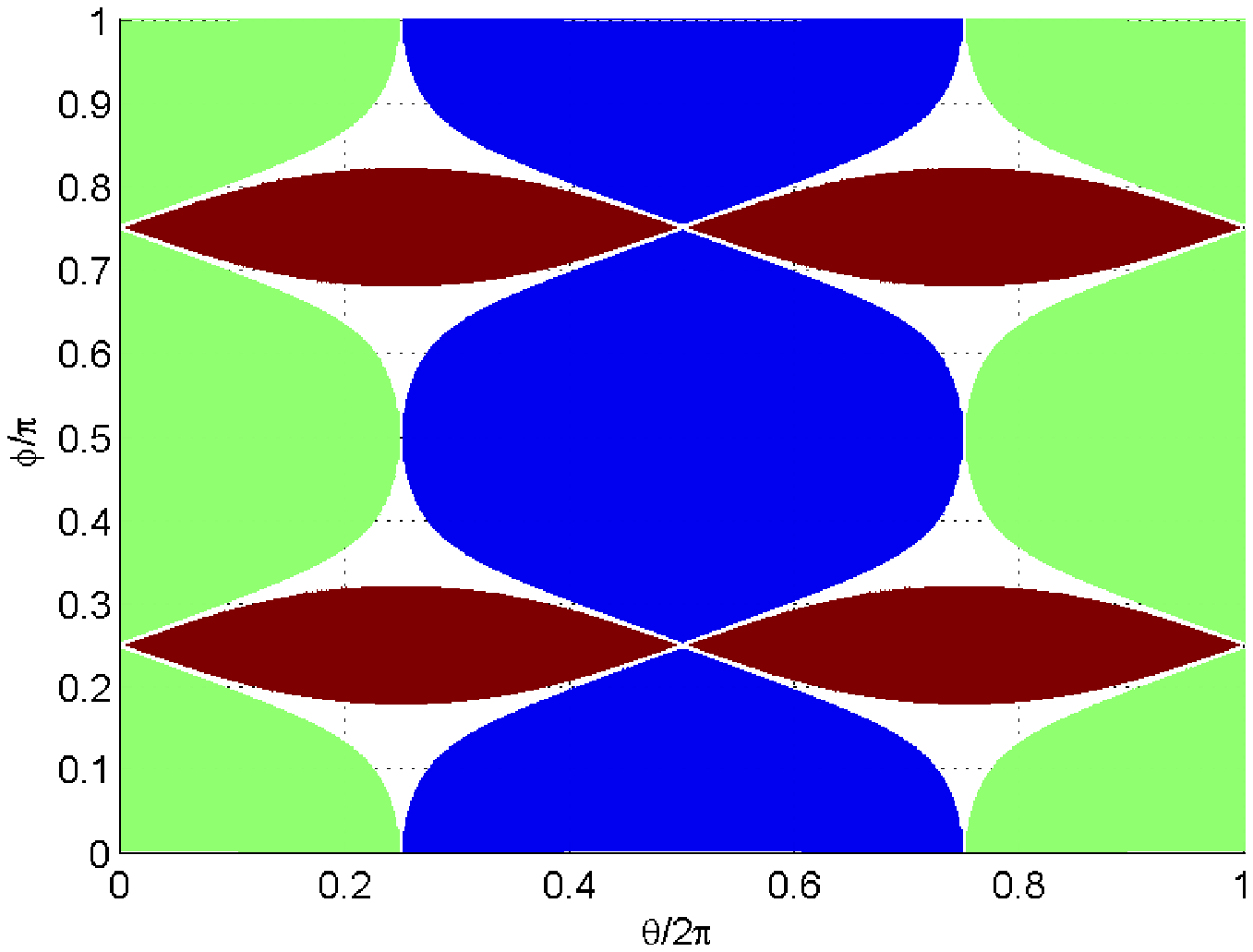}}
\end{center}
\caption{\small (Color online) The violations of the CHSH inequality with observables from the $x$-$z$ plane of Pauli sphere. Here, the red region is for $\rho_{(A_3)}$. The blue region is for $\rho_{(A_2)}$. The green region is for $\rho_{(A_1)}$. It shows that there is no choice of $\alpha, \beta$ and $\gamma$ such that the statistics of $\rho_{(A_j)}$'s violate the CHSH inequality.}
\label{fig-S2}
\end{figure}

Similarly, take local observable $A_1$ and $A_2$ for the first party, $C_1=\cos\theta_1{}X+i\sin\theta_1{}Y$ and $C_2=\cos\theta_1{}X-i\sin\theta_1{}Y$ for the third party. The statistics generated by local measurements on the state of $\rho_{(A_2)}$ violates the CHSH inequality \cite{CHSH} if $|\alpha\gamma|>\frac{1}{2\sqrt{2}}$, where $\theta_1=\frac{\pi}{4}$. Finally, by taking local observable $B_1'=Y$ and $B_2'=X$ for the second party, $C_1=\cos\theta_2{}X+i\sin\theta_2{}Y$ and $C_2=\cos\theta_2{}X-i\sin\theta_2{}Y$ for the third party, the statistics generated by local measurements on the state of $\rho_{(A_1)}$ violates the CHSH inequality \cite{CHSH} if $|\alpha\beta|>\frac{1}{2\sqrt{2}}$, where $\theta_2=\frac{\pi}{2}$. Now, the values of $\alpha, \beta$ and $\gamma$ which satisfy all the violation conditions are shown in Fig.\ref{fig-S3}. Hence, there is no choice of $\alpha, \beta$ and $\gamma$ such that the correlations derived from local measurements on $\rho_{(A_3)}$, $\rho_{(A_2)}$ and $\rho_{(A_1)}$ can violate the CHSH inequality \cite{CHSH} simultaneously. Here, each party can use different local observables dependent on the verified states \cite{Fei}.

Moreover, take observables from the $x$-$z$ plane of Pauli sphere, that is, Pauli matrices $X$ and $Z$. All the violations are shown in Fig.\ref{fig-S2} vias two phases. It shows that there is no choice of $\alpha, \beta$ and $\gamma$ such that the quantum correlations derived from local measurements on $\rho_{(A_i)}$'s violate the CHSH inequality \cite{CHSH}. This means that the CHSH inequality cannot be used for verifying the strong nonlocality of any W state in Eq.(E1). While the strong nonlocality of W states can be verified by violating the inequality (54), as shown in Fig. 9.

\section{Proof of the inequality (\ref{eqn5})}

Consider a separable state on Hilbert space $\mathbbm{H}_{A}\otimes\mathbbm{H}_B$ as
\begin{eqnarray}
\rho_{AB}=\sum_jp_j|\phi_j\rangle_A\langle \phi_j|\otimes|\varphi_j\rangle_B\langle \varphi_j|,
\label{L5}
\end{eqnarray}
where $|\phi_j\rangle=\alpha_{j0}|0\rangle+\alpha_{j1}|1\rangle$, $|\varphi_j\rangle=\beta_{j0}|0\rangle+\beta_{j1}|1\rangle$, and $\{p_j\}$ is a probability distribution. Let $\rho=(\rho_{s_1s_2;j_1j_2})$ be density matrix of $\rho$. We firstly prove that
\begin{eqnarray}
4\rho_{00;11}\rho_{11;00}-(\rho_{01;01}+\rho_{10;10})^2\leq 0.
\label{L6}
\end{eqnarray}
If $\rho_{AB}$ is given by a product state $|\phi_j\rangle_{A}|\varphi_j\rangle_{B}$, we have $4\rho_{00;11}\rho_{11;00}=4|\alpha_{j0}\beta_{j0}\alpha_{j1}\beta_{j1}|^2$, and $(\rho_{01;01}+\rho_{10;10})^2=(|\alpha_{j0}\beta_{j1}|^2+|\alpha_{j1}\beta_{j0}|^2)^2$. By using the Cauchy-Schwartz inequality of $2xy\leq x^2+y^2$, it follows the inequality (\ref{L6}). From the Hermitian symmetry of density matrix $\rho$ that $4\rho_{00;11}\rho_{11;00}=4|\rho_{00;11}|^2$. It follows that
\begin{eqnarray}
4\rho_{00;11}\rho_{11;00}&=&4|\rho_{00;11}|^2
\nonumber
\\
&=&4|\sum_jp_j\rho^{(j)}_{00;11}|^2
\label{L7}
\\
&\leq &(\sum_j2p_j|\rho^{(j)}_{00;11}|)^2
\label{L8}\\
&=&(\sum_j2p_j|\alpha_{j0}\beta_{j0}\alpha_{j1}\beta_{j1}|)^2
\nonumber\\
&\leq&(\sum_jp_j|\alpha_{j0}\beta_{j1}|^2+\sum_jp_j|\alpha_{j1}\beta_{j0}|^2)^2
\nonumber\\
\label{L9}
\\
&=&(\sum_jp_j(\rho_{01;01}^{(j)}+\rho_{10;10}^{(j)}))^2
\label{L10}
\\
&=&(\rho_{01;01}+\rho_{10;10})^2.
\label{L11}
\end{eqnarray}
Here, $\rho^{(j)}_{00;11}$ in Eq.(\ref{L7}) is given by $\rho^{(j)}_{00;11}=\alpha_{j0}\beta_{j0}\alpha_{j1}\beta_{j1}$. The inequality (\ref{L8}) follows from the triangle inequality of $|x+y|\leq |x|+|y|$.   The inequality (\ref{L9}) follows from the Cauchy-Schwartz inequality of $x^2+y^2\geq 2xy$.  $\rho_{01;01}^{(j)}$ and $\rho_{10;10}^{(j)}$ in Eq.(\ref{L10}) are given respectively by
$\rho_{01;01}^{(j)}=|\alpha_{j0}\beta_{j1}|^2$ and $\rho_{10;10}^{(j)}=|\alpha_{j1}\beta_{j0}|^2$. This proves the inequality (\ref{L6}).
\begin{figure}
\begin{center}
\resizebox{240pt}{220pt}{\includegraphics{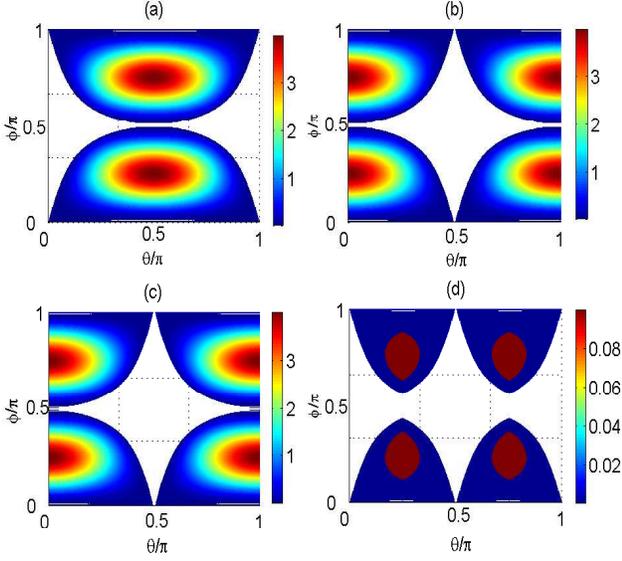}}
\end{center}
\caption{\small (Color online) The violations of the inequality (\ref{K4}) by measuring the states $\rho_{(A_1)}, \rho_{(A_2)}, \rho_{(A_3)}$ with observable from Pauli sphere.}
\label{fig-S4}
\end{figure}

Now, by using the definitions of Pauli matrices $X, Y, Z$, we get
$\rho_{01;01}=\frac{1}{4}\langle (\mathbbm{1}+Z)(\mathbbm{1}-Z)\rangle$, $\rho_{10;10}=
\frac{1}{4}\langle(\mathbbm{1}-Z)(\mathbbm{1}+Z)\rangle$, $\rho_{11;00}=-\frac{1}{4}\langle(iX-Y)(iX-Y)\rangle$ and $\rho_{00;11}=-\frac{1}{4}\langle(iX+Y)(iX+Y)\rangle$ with $i=\sqrt{-1}$. It follows the inequality (\ref{eqn5}).

In what follows, we construct a generalized Hardy-type inequality with three inputs \cite{Hardy}. Consider six distributions $\{P(ab|xy),a,b\in \{0,1\}\}$ for any $x,y\in \{0,1,2\}$. From the local model, it can be decomposed into
\begin{eqnarray}
P(ab|xy)&=&\int P(ab|xy,\lambda)d\mu(\lambda)
\nonumber
\\
&=&\int p(a|x,\lambda)p(b|y,\lambda)d\mu(\lambda),
\label{Ln1}
\end{eqnarray}
where $\lambda$ denotes any shared randomness with the measure $\mu(\lambda)$. Suppose that these distributions satisfy
\begin{eqnarray}
\sum_{a,b}(-1)^{a+b}P(ab|11)&=&\sum_{a,b}(-1)^{a+b}P(ab|12)
 \nonumber\\
&=&\sum_{a,b}(-1)^{a+b}P(ab|21)
 \nonumber\\
&=&\sum_{a,b}(-1)^{a+b}P(ab|22),
\label{Ln2a}
\end{eqnarray}
and
\begin{eqnarray}
&&P(01|00)P(10|00)=0,
\label{Ln2b}\\
&&P(00|11)=P(00|22),
\label{Ln2c}\\
&&P(10|11)\geq P(10|12),
\label{Ln2d}
\\
&&(P(01|11)-P(10|11))^2
\nonumber
\\
&&-(P(01|12)-P(10|12))^2
\nonumber\\
&& -(P(00|00)+P(11|00))^2\leq 0.
\label{Ln3}
\end{eqnarray}
Here, the constrictions (\ref{Ln2a})-(\ref{Ln2d}) are inspired by the joint conditional distributions associated from Pauli matrices on special separable states. In fact, from Eq.(\ref{Ln2a}), it follows that
\begin{eqnarray}
P(01|11)+P(10|11)=P(01|12)+P(10|12)
\end{eqnarray}
From the inequality (\ref{Ln2c}), it follows that
\begin{eqnarray}
(P(01|11)-P(10|11))^2\leq (P(01|12)-P(10|12))^2
\end{eqnarray}
Combined with Eq.(\ref{Ln2b}), it follows the inequality (\ref{Ln3}).

\section{Proof of the inequality (\ref{multinon})}

Consider a biseparable state on Hilbert space $\otimes_{j=1}^n\mathbbm{H}_{A_j}$ as
\begin{eqnarray}
\rho_{A_1\cdots {}A_n}=\sum_jp_j|\Phi_j\rangle_{S}\langle \Phi_j|\otimes|\Psi_j\rangle_{\overline{S}}\langle \Psi_j|,
\label{L15}
\end{eqnarray}
where $|\Phi_t\rangle=\sum_{j_1,\cdots,j_k=0}^{1}\alpha_{t,j_1\cdots{}j_k}|j_1\cdots{}j_k\rangle$ denotes the joint system shared by all parties in the subset $S=\{A_{j_1}, \cdots, A_{j_k}\}\subset\{A_1, \cdots, A_n\}$, $|\Psi_t\rangle=\sum_{s_1, \cdots, s_{n-k}=0}^{1}\beta_{t,s_1\cdots{}s_{n-k}}|j\rangle$ denotes the joint system shared by all the parties in the complement set $\overline{S}=\{A_{s_{1}}, \cdots, A_{s_{n-k}}\}$ of $S$, $s_t\not=j_\ell$ for any $t,\ell$, $\alpha_{t,j_1\cdots{}j_k}$ satisfies $\sum_{j_1,\cdots,j_k=0}^{1}|\alpha_{t,j_1\cdots{}j_k}|^2=1$, $\beta_{t,j_1\cdots{}j_k}$ satisfies $\sum_{s_1, \cdots, s_{n-k}=0}^{1}|\beta_{t,s_1\cdots{}s_{n-k}}|^2=1$, and $\{p_i\}$ is a probability distribution. Let $\rho=(\rho_{\vec{j}_1\vec{s}_2;\vec{j}'_1\vec{s}'_2})$ be the density matrix of $\rho$ with $\vec{j}_1=j_1\cdots{}j_k$ and $\vec{s}_2=s_1\cdots{}s_{n-k}$.

We firstly prove the following inequality
\begin{eqnarray}
4\rho_{\vec{0}_1\vec{0}_2;\vec{1}_1\vec{1}_2}\rho_{\vec{1}_1\vec{1}_2;\vec{0}_1\vec{0}_2}-
(1-\rho_{\vec{0}_1\vec{0}_2;\vec{0}_1\vec{0}_2}-\rho_{\vec{1}_1\vec{1}_2;\vec{1}_1\vec{1}_2})^2
\leq0.
\nonumber\\
\label{density}
\end{eqnarray}

If $\rho_{A_1\cdots{}A_n}$ is a product state given by $|\Phi_i\rangle_{S}|\Psi_i\rangle_{\overline{S}}$, we have
\begin{eqnarray}
4\rho_{\vec{0}_1\vec{0}_2;\vec{1}_1\vec{1}_2}\rho_{\vec{1}_1\vec{1}_2;\vec{0}_1\vec{0}_2}
&=&4|\alpha_{i,\vec{0}_1}\beta_{i,\vec{0}_2}\alpha_{i,\vec{1}_1}\beta_{i,\vec{1}_2}|^2
\nonumber
\\
&\leq& (|\alpha_{i,\vec{0}_1}\beta_{i,\vec{0}_2}|^2+|\alpha_{i,\vec{1}_1}\beta_{i,\vec{1}_2}|^2)^2
\nonumber
\\
&=&(\rho_{\vec{0}_1\vec{0}_2;\vec{0}_1\vec{0}_2}+\rho_{\vec{1}_1\vec{1}_2;\vec{1}_1\vec{1}_2})^2
\label{L17}
\end{eqnarray}
from the Cauchy-Schwartz inequality.

Now, consider the mixed state in Eq.(\ref{L15}). It follows from the inequality (\ref{L17}) that
\begin{eqnarray} &&4\rho_{\vec{0}_1\vec{0}_2;\vec{1}_1\vec{1}_2}\rho_{\vec{1}_1\vec{1}_2;\vec{0}_1\vec{0}_2}
\nonumber
\\&=&4|\rho_{\vec{0}_1\vec{0}_2;\vec{1}_1\vec{1}_2}|^2
\nonumber
\\
&=&4|\sum_jp_j\rho^{(j)}_{\vec{0}_1\vec{0}_2;\vec{1}_1\vec{1}_2}|^2
\label{L117}
\\
&\leq &(\sum_j2p_j|\rho^{(j)}_{\vec{0}_1\vec{0}_2;\vec{1}_1\vec{1}_2}|)^2
\label{L118}\\
&=&(\sum_j2p_j|\alpha_{j,\vec{0}_1}\beta_{j,\vec{0}_2}\alpha_{j,\vec{1}_1}\beta_{j,\vec{1}_2}|)^2
\nonumber\\
&\leq&(\sum_jp_j|\alpha_{j,\vec{0}_1}\beta_{j,\vec{1}_2}|^2
+\sum_jp_j|\alpha_{j,\vec{1}_1}\beta_{j,\vec{0}_2}|^2)^2
\label{L119}
\\
&=&(\sum_jp_j(\rho_{\vec{0}_1\vec{1}_2;\vec{1}_0\vec{0}_2}^{(j)}
+\rho_{\vec{1}_1\vec{0}_2;\vec{1}_1\vec{0}_2}^{(j)}))^2
\nonumber
\\
&\leq &(\sum_{\vec{j}_1\vec{s}_2\not\in\{\vec{0}_1\vec{0}_2;\vec{1}_1\vec{1}_2\}}
\rho_{\vec{j}_1\vec{s}_2;\vec{j}_1\vec{s}_2})^2
\label{L21}
\\
&= &(1-\rho_{\vec{0}_1\vec{0}_2;\vec{0}_1\vec{0}_2}-\rho_{\vec{1}_1\vec{1}_2;\vec{1}_1\vec{1}_2})^2.
\label{L22}
\end{eqnarray}
Here, the inequality (\ref{L117}) follows from the triangle inequality of $|x+y|\leq |x|+|y|$. $\rho^{(j)}_{\vec{0}_1\vec{0}_2;\vec{1}_1\vec{1}_2}$ in Eq.(\ref{L118}) is given by $\rho^{(j)}_{\vec{0}_1\vec{0}_2;\vec{1}_1\vec{1}_2}=
\alpha_{j,\vec{0}_1}\beta_{j,\vec{0}_2}\alpha_{j,\vec{1}_1}\beta_{j,\vec{1}_2}$. The inequality (\ref{L119}) is followed from the Cauchy-Schwartz inequality of $x^2+y^2\geq 2xy$. The inequality (\ref{L21}) is obtained by using the inequality (\ref{L17}) for all the bipartitions of $S$ and $\overline{S}$. The equality (\ref{L22}) is followed from the trace equality of ${\rm tr}\rho=1$. This completes the proof of the inequality (\ref{density}).

Now, using the definitions of Pauli matrices $X, Y, Z$, we get
\begin{eqnarray*}
&&\rho_{\vec{0}_1\vec{0}_2;\vec{0}_1\vec{0}_2}=\frac{1}{2^n}\langle (\mathbbm{1}+Z)^{\otimes n}\rangle,
 \\
&&\rho_{\vec{1}_1\vec{1}_2;\vec{1}_1\vec{1}_2}=\frac{1}{2^n}\langle (\mathbbm{1}-Z)^{\otimes n}\rangle,
\\
&&\rho_{\vec{0}_1\vec{0}_2;\vec{1}_1\vec{1}_2}=
-\frac{1}{2^n}\langle(iX+Y)^{\otimes n}\rangle,
\\
&&\rho_{\vec{1}_1\vec{1}_2;\vec{0}_1\vec{0}_2}=
-\frac{1}{2^n}\langle(iX-Y)^{\otimes n}\rangle
\end{eqnarray*}
with $i=\sqrt{-1}$. It follows the inequality (\ref{multinon}).

Moreover, the inequality (\ref{multinon}) may be used to construct Hardy-type inequality similar to the inequalities (\ref{Ln2a})-(\ref{Ln3}).

Now, we consider a biseparable state on $d$-dimensional Hilbert space $\otimes_{j=1}^n\mathbbm{H}_{A_j}$ with $d\geq 2$ as
\begin{eqnarray}
\rho_{A_1\cdots{}A_n}=\sum_jp_j|\Phi_j\rangle_{S}\langle \Phi_j|\otimes|\Psi_j\rangle_{\overline{S}}\langle \Psi_j|,
\label{Lh15}
\end{eqnarray}
where
\begin{eqnarray}
|\Phi_t\rangle=\sum_{j_1,\cdots,j_k=0}^{d-1}\alpha_{t,j_1\cdots{}j_k}|j_1\cdots{}j_k\rangle
\end{eqnarray}
denotes the joint system shared by all parties in the subset $S=\{A_{j_1}, \cdots, A_{j_k}\}\subset\{A_1, \cdots, A_n\}$, the state of
\begin{eqnarray}
|\Psi_t\rangle=\sum_{s_1, \cdots, s_{n-k}=0}^{d-1}\beta_{t,s_1\cdots{}s_{n-k}}|j\rangle
\end{eqnarray}
denotes the joint system shared by all the parties in the complement set $\overline{S}=\{A_{s_{1}}, \cdots, A_{s_{n-k}}\}$ of $S$, $s_t\not=j_\ell$ for any $t,\ell$, $\alpha_{t,j_1\cdots{}j_k}$ satisfies $\sum_{j_1,\cdots,j_k=0}^{d-1}|\alpha_{t,j_1\cdots{}j_k}|^2=1$, $\beta_{t,j_1\cdots{}j_k}$ satisfies $\sum_{s_1, \cdots, s_{n-k}=0}^{d-1}|\beta_{t,s_1\cdots{}s_{n-k}}|^2=1$, and $\{p_i\}$ is a probability distribution. Let
\begin{eqnarray}
\rho=(\rho_{\vec{j}_1\vec{s}_2;\vec{j}'_1\vec{s}'_2})
\end{eqnarray}
be the density matrix of $\rho$ with $\vec{j}_1=j_1\cdots{}j_k$ and $\vec{s}_2=s_1\cdots{}s_{n-k}$.

We firstly prove the following inequality
\begin{eqnarray}
&&\!\!\!\!4\rho_{\vec{0}_1\vec{0}_2;\overrightarrow{d-1}_1\overrightarrow{d-1}_2}
\rho_{\overrightarrow{d-1}_1\overrightarrow{d-1}_2;\vec{0}_1\vec{0}_2}
\nonumber
\\
&&\!\!\!\!-
(1-\rho_{\vec{0}_1\vec{0}_2;\vec{0}_1\vec{0}_2}-\rho_{\overrightarrow{d-1}_1\overrightarrow{d-1}_2;
\overrightarrow{d-1}_1\overrightarrow{d-1}_2})^2\leq0
\label{hdensity}
\end{eqnarray}

If $\rho_{A_1\cdots{}A_n}$ is a product state given by $|\Phi_i\rangle_{S}|\Psi_i\rangle_{\overline{S}}$, we have
\begin{eqnarray}
&&4\rho_{\vec{0}_1\vec{0}_2;\overrightarrow{d-1}_1\overrightarrow{d-1}_2}
\rho_{\overrightarrow{d-1}_1\overrightarrow{d-1}_2;\vec{0}_1\vec{0}_2}
\nonumber
\\
&=&4|\alpha_{i,\vec{0}_1}\beta_{i,\vec{0}_2}
\alpha_{i,\overrightarrow{d-1}_1}\beta_{i,\overrightarrow{d-1}_2}|^2
\nonumber
\\
&\leq& (|\alpha_{i,\vec{0}_1}\beta_{i,\vec{0}_2}|^2+|\alpha_{i,\overrightarrow{d-1}_1}
\beta_{i,\overrightarrow{d-1}_2}|^2)^2
\nonumber
\\
&=&(\rho_{\vec{0}_1\vec{0}_2;\vec{0}_1\vec{0}_2}
+\rho_{\overrightarrow{d-1}_1\overrightarrow{d-1}_2;\overrightarrow{d-1}_1
\overrightarrow{d-1}_2})^2
\label{Lh17}
\end{eqnarray}
from the Cauchy-Schwartz inequality.

Now, consider the mixed state in Eq.(\ref{Lh15}). It follows from the inequality (\ref{Lh17}) that
\begin{eqnarray}
&&4\rho_{\vec{0}_1\vec{0}_2;\overrightarrow{d-1}_1\overrightarrow{d-1}_2}
\rho_{\overrightarrow{d-1}_1\overrightarrow{d-1}_2;\vec{0}_1\vec{0}_2}
\nonumber
\\
&=&4|\rho_{\vec{0}_1\vec{0}_2;\overrightarrow{d-1}_1\overrightarrow{d-1}_2}|^2
\nonumber
\\
&=&4|\sum_jp_j\rho^{(j)}_{\vec{0}_1\vec{0}_2;\overrightarrow{d-1}_1\overrightarrow{d-1}_2}|^2
\label{Lh17}
\\
&\leq &(\sum_j2p_j|\rho^{(j)}_{\vec{0}_1\vec{0}_2;\overrightarrow{d-1}_1\overrightarrow{d-1}_2}|)^2
\label{Lh18}\\
&=&(\sum_j2p_j|\alpha_{j,\vec{0}_1}\beta_{j,\vec{0}_2}\alpha_{j,\overrightarrow{d-1}_1}
\beta_{j,\overrightarrow{d-1}_2}|)^2
\nonumber\\
&\leq&(\sum_jp_j|\alpha_{j,\vec{0}_1}\beta_{j,\overrightarrow{d-1}_2}|^2
+\sum_jp_j|\alpha_{j,\overrightarrow{d-1}_1}\beta_{j,\vec{0}_2}|^2)^2
\nonumber\\
\label{Lh19}
\\
&=&(\sum_jp_j(\rho_{\vec{0}_1\overrightarrow{d-1}_2;\overrightarrow{d-1}_0\vec{0}_2}^{(j)}
+\rho_{\overrightarrow{d-1}_1\vec{0}_2;\overrightarrow{d-1}_1\vec{0}_2}^{(j)}))^2
\nonumber
\\
&\leq &(\sum_{\vec{j}_1\vec{s}_2\not\in\{\vec{0}_1\vec{0}_2;
\overrightarrow{d-1}_1\overrightarrow{d-1}_2\}\atop{j_t,s_\ell\in \{0,d-1\}}}
\rho_{\vec{j}_1\vec{s}_2;\vec{j}_1\vec{s}_2})^2.
\label{Lh21}
\end{eqnarray}
Here, the inequality (\ref{Lh17}) follows from the triangle inequality of $|x+y|\leq |x|+|y|$. $\rho^{(j)}_{\vec{0}_1\vec{0}_2;\overrightarrow{d-1}_1\overrightarrow{d-1}_2}$ in Eq.(\ref{Lh18}) is given by $\rho^{(j)}_{\vec{0}_1\vec{0}_2;\overrightarrow{d-1}_1\overrightarrow{d-1}_2}=
\alpha_{j,\vec{0}_1}\beta_{j,\vec{0}_2}\alpha_{j,\overrightarrow{d-1}_1}\beta_{j,\vec{1}_2}$. The inequality (\ref{Lh19}) is followed from the Cauchy-Schwartz inequality of $x^2+y^2\geq 2xy$. The inequality (\ref{Lh21}) is obtained by using the inequality (\ref{Lh17}) for all the bipartitions of $S$ and $\overline{S}$.

Similarly, we get
\begin{eqnarray}
&&4\rho_{\vec{u}_1\vec{u}_2;\overrightarrow{d-u-1}_1\overrightarrow{d-u-1}_2}
\rho_{\overrightarrow{d-u-1}_1\overrightarrow{d-u-1}_2;\vec{u}_1\vec{u}_2}
\nonumber
\\
&\leq& (\sum_{\vec{j}_1\vec{s}_2\not\in\{\vec{u}_1\vec{u}_2;
\overrightarrow{d-u-1}_1\overrightarrow{d-u}_2\}\atop{j_t,s_\ell\in \{u,d-u-1\}}}
\rho_{\vec{j}_1\vec{s}_2;\vec{j}_1\vec{s}_2})^2
\label{Lh22}
\end{eqnarray}
for any $u<\frac{n}{2}$. From the inequalities (\ref{Lh21}) and (\ref{Lh22}), it follows that
\begin{eqnarray}
&&4\sum_{u=0}^{\lfloor\frac{d-1}{2}\rfloor}
(\rho_{\vec{u}_1\vec{u}_2;\overrightarrow{d-u-1}_1\overrightarrow{d-u-1}_2}
\rho_{\overrightarrow{d-u-1}_1\overrightarrow{d-u-1}_2;\vec{u}_1\vec{u}_2})^{1/2}
\nonumber
\\
&\leq & \sum_{u=0}^{\lfloor\frac{d-1}{2}\rfloor}\sum_{\vec{j}_1\vec{s}_2\not\in\{\vec{u}_1\vec{u}_2;
\overrightarrow{d-u-1}_1\overrightarrow{d-u}_2\}\atop{j_t,s_\ell\in \{u,d-u-1\}}}
\rho_{\vec{j}_1\vec{s}_2;\vec{j}_1\vec{s}_2}
\\
&\leq & \sum_{u=0}^{d-1}\rho_{\vec{u}_1\vec{u}_2;\vec{d-1-u}_1\vec{d-1-u}_2}
\label{Lh23}
\end{eqnarray}
from the trace equality of ${\rm Tr}\rho=1$. This completes the proof of the inequality (\ref{hdensity}). In experiment, we can take use of generalized Gell-Mann matrices \cite{Loub}.

\section{The star-convexity of ${\cal B}_d^{(k)}$}

The proof is similar to its shown in Appendix D. Consider Hilbert space $\mathbb{H}:=\otimes_{j=1}^n\mathbb{H}_{A_j}$. The goal is to construct one state $\rho=p\rho_1+(1-p)\rho_2$ such that $\rho\not\in {\cal B}_d^{(k)}$ for some $p>0$ and $\rho_1,\rho_2\in {\cal B}_d^{(k)}$. Define
\begin{eqnarray}
\rho_1&=&|\Phi_1\rangle_{A_1\cdots{}A_4} \langle\Phi_1|,
\nonumber\\
\rho_2&=&|\Phi_1\rangle_{A_1\cdots{}A_4} \langle\Phi_2|
\label{eqna1}
\end{eqnarray}
with
\begin{eqnarray}
|\Phi_1\rangle=\frac{1}{2}(|0001\rangle+|0010\rangle+|0100\rangle+|1001\rangle),
\nonumber\\
|\Phi_2\rangle=\frac{1}{2}(|1000\rangle+|0010\rangle+|0100\rangle+|1001\rangle).
\label{eqna2}
\end{eqnarray}
It is easy to prove that $|\Phi_1\rangle$ and $|\Phi_2\rangle$ are genuinely 4-partite entangled states. This can be easily completed by using the Schmidt decomposition for each bipartition. Moreover, denote $\rho_{(A_i)}^{(j)}$ as the output state of $\rho_j$ after its passing through the particle-lose channel ${\cal E}_{A_i}(\cdot)$, $i=1, \cdots, 4; j=1, 2$. The density matrices of $\rho_{(A_i)}^{(j)}$ are given by
\begin{eqnarray}
\rho_{(A_1)}^{(1)}&=&\rho_{(A_4)}^{(2)}
=\frac{3}{4}|W_1\rangle\langle W_1|+\frac{1}{4}|001\rangle\langle 001|,
\label{eqna3}\\
\rho_{(A_2)}^{(1)}&=&\rho_{(A_3)}^{(1)}=\rho_{(A_2)}^{(2)}=\rho_{(A_3)}^{(2)}
\nonumber
\\
&=&\frac{3}{4}|W_2\rangle\langle W_2|+\frac{1}{4}|000\rangle\langle 000|,
\label{eqna4}\\
\rho_{(A_4)}^{(1)}&=&\rho_{(A_1)}^{(2)}=\frac{1}{2}|+\rangle\langle +|\otimes|00\rangle\langle 00|
\nonumber
\\&&+ \frac{1}{2}|0\rangle\langle 0|\otimes|\phi_{01}\rangle\langle \phi_{01}|,
\label{eqna5}
\end{eqnarray}
where $|W_j\rangle$'s are given by
\begin{eqnarray}
|W_1\rangle&=&\frac{1}{\sqrt{3}}(|001\rangle+|010\rangle+|101\rangle),
\\
|W_2\rangle&=&\frac{1}{\sqrt{3}}(|001\rangle+|010\rangle+|100\rangle),
\end{eqnarray}
$|+\rangle=\frac{1}{\sqrt{2}}(|0\rangle+|1\rangle)$
and $|\phi_{01}\rangle=\frac{1}{\sqrt{2}}(|01\rangle+|10\rangle)$. $\rho_{(A_1)}^{(1)}, \rho_{(A_2)}^{(1)}, \rho_{(A_3)}^{(1)}, \rho_{(A_2)}^{(2)}, \rho_{(A_3)}^{(2)}$ and $\rho_{(A_4)}^{(2)}$ are genuinely tripartite entangled states \cite{Sy} because $|W_1\rangle$ and $|W_2\rangle$ are genuinely tripartite entangled states \cite{DVC}. Moreover, $\rho_4^{(1)}$ and $\rho_1^{(2)}$ are biseparable states. It means that $\rho_1, \rho_2\in {\cal B}_d^{(3)}$.

In what follows, define $\varrho=\frac{1}{2}\rho_1+\frac{1}{2}\rho_2$. From Eqs.(\ref{eqna3})-(\ref{eqna5}), it follows that
\begin{eqnarray}
\varrho_{(A_1)}^{(1)}&=&\varrho_{(A_4)}^{(1)}
= \frac{3}{8}|W_1\rangle\langle W_1|+\frac{1}{4}|+\rangle\langle+|\otimes|00\rangle\langle 00|
\nonumber
\\&&+\frac{1}{8}|001\rangle\langle 001|
+ \frac{1}{4}|0\rangle\langle 0|\otimes|\phi_{01}\rangle\langle \phi_{01}|,
\label{eqna6}\\
\varrho_{(A_2)}^{(1)}&=&\varrho_{(A_3)}^{(1)}
=\frac{3}{4}|W_2\rangle\langle W_2|+\frac{1}{4}|000\rangle\langle 000|.
\label{eqna7}
\end{eqnarray}
Note that $\varrho$ is a genuinely $4$-partite entanglement \cite{Sy} since $|\Phi_1\rangle$ and $|\Phi_2\rangle$ are genuinely $4$-partite entangled states. All the final states $\varrho_{(A_i)}={\cal E}_{A_i}(\varrho)$ are also genuinely tripartite entangled states \cite{Sy}. It means that $\varrho\not\in {\cal B}_d^{(3)}$. This implies that ${\cal B}_d^{(3)}$ is not convex.

The star convexity of ${\cal B}_d^{(3)}$ is followed by choosing the maximally mixed state $\rho_{0}=\frac{1}{8}\mathbbm{1}$ as the center point. For any state $\rho\in {\cal B}_d^{(3)}$, it is easy to prove that $p\rho+(1-p)\rho_0\in {\cal B}_d^{(3)}$ for any $p\in (0,1)$. Similar result holds for ${\cal B}_d^{(k)}$.

\section{Proof of Lemma 4}

Consider any pair of two parties $\textsf{A}_i$ and $\textsf{A}_j$ with $1\leq i<j\leq n$. Our goal is to prove that there exist $k$ chain-type subnetworks (edge-disjoint-paths) ${\cal L}^{(i,j)}_1, \cdots,  {\cal L}^{(i,j)}_k$, where ${\cal L}^{(i,j)}_s=\{\textsf{A}_i, \textsf{A}_{s_1}, \cdots, \textsf{A}_{s_t}, \textsf{A}_j\}$ consists a long chain-type network satisfying ${\cal L}^{(i,j)}_s$ and ${\cal L}^{(i,j)}_\ell$ have at most one common party or multiple parties who are not adjacent (or sharing bipartite entanglement) for any $s\not=\ell$. This can be proved by induction on the number of parties.

Consider the graphic representation ${\cal N}$ of quantum network ${\cal N}_q$, where each party is schematically denoted as one node and each bipartite entangled pure state shared by two parties is denoted as one edge linked to two nodes. For each pair of $\textsf{A}_i$ and $\textsf{A}_{j}$ who are connected by one edge, from Menger Theorem \cite{Menger}, the minimum size of the edge-separator equals the maximum number of pairwise edge-disjoint-paths. This means that there are at least $k$ pairwise edge-disjoint-paths connecting $\textsf{A}_i$ and $\textsf{A}_{j}$. Otherwise, there are only $k-1$ edge-disjoint-paths ${\cal L}^{(i,j)}_1, \cdots,  {\cal L}^{(i,j)}_{k-1}$, that is, there is one edge $e$ from $\textsf{A}_i$ that cannot result in one path connecting to $\textsf{A}_{j}$. However, for the edge $e$, one always finds another path ${\cal L}^{(i,j)}_k$ connecting two parties $\textsf{A}_i$ and $\textsf{A}_{i+1}$, which are different from other paths of ${\cal L}^{(i,j)}_1, \cdots,  {\cal L}^{(i,j)}_{k-1}$. The main reason is that each node will be used only once in one path. Otherwise, there is a cycle which should be deleted. Hence, from the vertex of $e$, one can always find a new edge $e_1$ which has not been used in ${\cal L}^{(i,j)}_1, \cdots,  {\cal L}^{(i,j)}_{k-1}$.

This procedure can be iteratively forward. In each time, one new party and one connected edge $e_t$ will be added into ${\cal L}^{(i,j)}_k$. Since there are only $n$ parties, the iteration will be ended with the party $\textsf{A}_{j}$. It means that there is another path ${\cal L}^{(i,j)}_k$ for the edge $e$. This contradicts to the assumption that there are only $k-1$ edge-disjoint-paths ${\cal L}^{(i,j)}_1, \cdots,  {\cal L}^{(i,j)}_{k-1}$. So, there are $k$ edge-disjoint-paths for $\textsf{A}_i$ and $\textsf{A}_{j}$. This proves Lemma 4.

\end{document}